\documentclass[useAMS,usenatbib]{mn2e}
\usepackage{graphicx}
\usepackage{epstopdf}
\usepackage{rotating}
\usepackage{aas_macros}

\newcommand{\LCDM}{$\Lambda$CDM}
\newcommand{\msun}{\mbox{${\rm M}_{\odot}$}}
\newcommand{\msunpcsq}{\mbox{${\rm M}_{\odot} {\rm pc}^{-2}$}}
\newcommand{\zsun}{\mbox{${\rm Z}_{\odot}$}}

\newcommand{\HII}{\mbox{${\rm H}_{\rm II}$}}
\newcommand{\HI}{\mbox{${\rm H}_{\rm I}$}}
\newcommand{\Htwo}{\mbox{${\rm H}_{2}$}}
\newcommand{\NHI}{\mbox{${N}_{\rm HI}$}}

\newcommand{\atomscm}{\mbox{${\rm atoms ~ cm}^{-2}$}}
\newcommand{\msunpc}{\mbox{${\rm M}_\odot {\rm pc}^{-2}$}}
\newcommand{\sDLA}{\mbox{${\sigma}_{\rm DLA}$}}

\def\lesssim{\lower.5ex\hbox{$\; \buildrel < \over \sim \;$}}
\def\gtrsim{\lower.5ex\hbox{$\; \buildrel > \over \sim \;$}}

\title[DLAs in SAMs with Multiphase Gas]{Damped Ly$\alpha$ Absorption Systems in Semi-Analytic Models with Multiphase Gas}

\author[Berry et al.]{Michael Berry$^{1}$, 
Rachel S. Somerville$^{1}$, 
Marcel R. Haas$^{1}$, 
Eric Gawiser$^{1}$, \newauthor 
Ari Maller$^{2,~3}$, 
Gerg\"o Popping$^{4}$, 
Scott C. Trager$^{4}$ \\
$^{1}$Department of Physics and Astronomy, Rutgers University, Piscataway, NJ 08854-8019, U.S.A. \\
$^{2}$Physics Department, New York City College of Technology, Brooklyn, NY 11201, USA \\
$^{3}$Department of Astrophysics, American Museum of Natural History, New York, NY 10024, USA \\
$^{4}$Kapteyn Astronomical Institute, University of Groningen, Postbus 800, NL-9700 AV Groningen, the Netherlands}

\begin{document}

\pagerange{\pageref{firstpage}--\pageref{lastpage}} \pubyear{2013}

\maketitle

\label{firstpage}

\begin{abstract}

We investigate the properties of damped Ly$\alpha$ absorption systems 
(DLAs) in semi-analytic models of galaxy formation, including new
modeling of the partitioning of cold gas into
atomic, molecular, and ionized phases, and a star formation recipe based on the density 
of molecular gas. We use three approaches 
for partitioning gas into atomic and molecular constituents: a
pressure-based recipe and metallicity-based recipes with fixed and 
varying UV radiation fields. 
We identify DLAs by adopting an 
assumed gas density profile for galactic discs and passing lines of
sight through our simulations to compute \HI\, column densities. We find
that models with ``standard'' gas radial profiles --- computed
assuming that the average specific angular momentum of the gas disc is
equal to that of the host dark matter halo --- fail to reproduce the
observed column density distribution of DLAs, regardless of the
assumed gas partitioning.  These models also
fail to reproduce the distribution of velocity widths $\Delta v$ of
low-ionization state metal systems, overproducing low $\Delta v$
relative to high $\Delta v$ systems.
Models with ``extended'' radial gas profiles --- corresponding to gas
discs with higher specific angular momentum, or gas in an alternate 
extended configuration --- are able to reproduce quite well
the column density distribution of absorbers over the column density
range $19 < \log~\NHI\ < 22.5$ in the redshift range $2 <  z < 3.5$. 
The model with pressure-based gas partitioning and the metallicity-based 
recipe with a varying UV 
radiation field also reproduce the observed line density of DLAs, \HI\, 
gas density, and $\Delta v$ distribution at $z<3$ well. 
However \emph{all} of the models investigated here underproduce DLAs 
and the \HI\, gas density at $z>3$. This may indicate that DLAs at high 
redshift arise from a different physical phenomenon, such as outflows 
or filaments.
If this is the case, the flatness in the number of DLAs and \HI\, gas 
density over the redshift 
interval $0<z<5$ may be due to a cosmic coincidence where the majority of 
DLAs at $z>3$ arise from intergalactic gas in filaments or streams 
while those at $z<3$ arise predominantly in galactic discs. 
We further investigate the dependence of DLA metallicity
on redshift and $\Delta v$ in our favored models, and find good agreement with
the observations, particularly when we include the effects of
metallicity gradients. 
\end{abstract}

\section{Introduction}
\label{Sec:Intro}

The study of gas in absorption against background quasars (quasar
absorption systems) has a long and rich observational history 
\citep{Wolfe1986, Wolfe1995, Storrie1996, Storrie2000, Peroux2003}. 
Damped Lyman-$\alpha$ systems (DLAs),
defined as systems with column density ${\rm log} ~\NHI\ > 20.3$ \atomscm,
are of particular interest for studies of galaxy formation because
they are believed to arise predominantly from neutral gas within or
closely associated with galaxies. DLAs are believed to contain the
majority of neutral gas in the Universe \citep{Storrie2000, Wolfe2000, Peroux2003, Prochaska2005, Noterdaeme2009} 
and are therefore 
reservoirs for future star formation. Indeed, absorption studies are 
currently the only means of probing the \HI\ content of galaxies at 
significant redshifts and provide an orthogonal means of studying 
galaxy evolution.
Observational properties that can be measured for DLAs
include their number density and column density distribution,
metallicities, and kinematic properties \citep{Lanzetta1995, Storrie1996, Pettini1994, Pettini1997, Prochaska1997, Prochaska2000}. These
observations provide important constraints on the gas content of
galaxies at high redshift, and indirectly constrain how gas is converted into
stars, and how gas and metals are cycled into and out of galaxies in
inflows and outflows. These, in turn, provide key constraints on some
of the most uncertain aspects of our models of galaxy formation.

There has been a significant amount of observational activity and
progress in this area in recent years. Large surveys such as the Sloan
Digital Sky Survey \citep[SDSS][]{Schneider2010} and BOSS 
\citep{Eisenstein2011} have provided extensive target samples of optically
detected quasars, yielding greatly improved statistics for samples of
high redshift absorbers ($1.5 \lesssim z \lesssim 4.5$). These improved
statistics have greatly tightened the constraints on the shape of the
column density distribution function, comoving line density of DLAs,
and the evolution of the cosmological neutral gas density 
\citep[e.g.][]{Noterdaeme2012}. 
\cite{Font-Ribera2012} employed a cross-correlation analysis
of DLAs from the BOSS survey with the Ly$\alpha$ forest and were able
to obtain constraints on the DLA cross-section as a function of halo
mass. \cite{Rafelski2012} and \cite{Neeleman2013}have published metallicities for a large
number of DLAs in the redshift interval $2<z<4$, providing constraints
on the build-up of heavy elements in the cold gas phase of galaxies
across cosmic time. In addition, the UV-sensitive Cosmic Origins
Spectrograph (COS) on the Hubble Space Telescope (HST) is now enabling
studies of DLAs at lower redshift $z\lesssim 1$, which may be more
easily connected with populations detected in emission and with the
present day galaxy population \citep{Meiring2011, Battisti2012}.
Recent studies with COS have also yielded a wealth of information on 
ionized gas within low-redshift haloes \citep[e.g.][]{Tumlinson2011}.

Two different pictures for the origin and nature of DLAs have been
debated in the literature.  Based on the observed kinematics of
low-ionization metal systems in DLAs, \cite{Wolfe1986} and
\citet[][hereafter PW97]{Prochaska1997} presented a picture in which
thick, extended disc galaxies give rise to DLAs, yet explaining how a
sufficient number of large disc galaxies could have formed by $z \sim
3$ remains a challenge.  In contrast, in the context of the
hierarchical picture arising in a Cold Dark Matter (CDM) cosmogony,
many DLAs would be expected to be associated with smaller, lower mass
systems \citep[e.g.][]{Haehnelt1998}. However, reproducing the
distribution of DLA kinematics has remained a significant challenge
for hierarchical models \citep{Maller2001,Razoumov2008,Pontzen2008}.
These two scenarios have very different implications for galaxy
evolution --- the former requires large disc galaxies to be in place
by $z \sim 3$, and the latter has implications for the expected star
formation rates, stellar masses, and kinematics of DLAs and their
counterparts.

A number of previous theoretical studies have made specific
predictions for the properties of DLAs in the framework of the CDM 
paradigm \citep[e.g.,][]{Kauffmann1994, Kauffmann1996, Gardner1997, Haehnelt1998, Maller2001, Maller2003, Nagamine2004_ZSFR, Nagamine2004_abund, Nagamine2007, Pontzen2008, Fumagalli2011, Altay2011, Cen2012, vandeVoort2012, Kulkarni2013, Altay2013}. 
Early numerical hydrodynamic simulations typically neglected feedback
from stellar and supernova-driven winds, or contained weak forms of
stellar feedback. These simulations had moderate success in
reproducing the column density distribution, cosmological neutral
gas density, and line density of DLAs at high redshift ($1 \lesssim z
\lesssim 4$), although they had difficulty reproducing the turnover in
the column density distribution at ${\rm log} ~ \NHI\ \sim 22$
\atomscm\ \citep{Nagamine2004_abund}. These simulations were not able
to discriminate between different phases of gas, and it was speculated
that the turnover could be due to the \HI-H$_2$ transition. However,
recent observational results from the BOSS survey
\citep{Font-Ribera2012} have shown that while these high column
density systems are rare, they do exist, relaxing some of this
tension, although the origin of the turnover still remains something
of a puzzle \citep{Erkal2012}. Many of these simulations had relatively
small volumes and modeled the DLA column density distribution by
characterizing the relationship between DLA cross-section and dark
matter (DM) halo mass, then convolving this relationship with a DM
halo mass function from larger volume dissipationless simulations.

More recent simulations found that the inclusion of more effective
stellar feedback and winds had a significant effect on the predicted
DM halo mass to DLA cross section relationship. For example,
\citet{Nagamine2007} found that in simulations with strong winds,
galaxies in low mass haloes ejected much of their gas, resulting in a
lower DLA cross section, thus shifting the DLA population into higher
mass host haloes. Qualitatively similar results have been found by
\citet{Pontzen2008}, \citet{Fumagalli2011}, and \citet{Cen2012},
although the detailed slope, normalization, and redshift dependence of
the predicted halo mass to DLA cross section relationship are
different in these different simulations. Recently, \cite{Cen2012}
particularly emphasized the importance of outflows for reproducing the
observed properties of DLAs including their kinematics.

CDM-based models have also had difficulty reproducing the observed
metallicities of DLAs \citep{Somerville2001, Maller2001, Nagamine2004_ZSFR, Nagamine2005, Pontzen2008, Fumagalli2011}. 
They have consistently
predicted higher average metallicities, and none have reproduced the
tail to very low metallicity, although again, simulations with strong
stellar feedback and winds have been more successful. \cite{Cen2012}
find that a significant number of $z \gtrsim 3$ DLAs originiate in
intergalactic gas. A combination of these intergalactic DLAs and the
ejection of metals by galactic winds lowers the average metallicities,
bringing them into better agreement with observations.

Semi-analytic models (SAMs), based within the framework of the CDM 
paradigm for structure formation \citep{Blumenthal1984}, 
have been widely used to produce a general picture of how
density fluctuations in the primordial universe evolve into the
observable galaxy population \citep{Percival2002, Tegmark2004,
Eisenstein2005}. Rather than solving detailed equations of hydrodynamics and
thermodynamics for individual particles or grid cells, SAMs use simple
but physically motivated ``recipes'' to track bulk quantities such as
the total mass in stars, hot gas, cold gas, metals, etc, in various
``zones'' (e.g. within a galaxy, within a dark matter halo, in a halo
infall region, or in the intergalactic medium). In some cases, SAMs
attempt to track these quantities in radial bins within a galactic
disc \citep{Kauffmann1996,Avila-Reese2001, Dutton2009, Fu2010, Kauffmann2012}. 
Although they cannot offer the detailed spatial and kinematic information 
provided by fully numerical hydrodynamical simulations,
SAMs do have a number of advantages over these techniques. Numerical
hydro simulations of galaxy formation still must rely heavily on
``sub-grid'' recipes for important processes such as star formation
and stellar feedback. These are treated in a similar manner in SAMs,
but the effect of varying the details of these recipes and their
parameters can be explored much more thoroughly because of their
greater computational efficiency. SAMs can provide ``mock catalogs''
for very large numbers of galaxies, comparable to modern surveys,
while this is still inaccessable for numerical hydro
simulations. Finally, we still do not understand many of the details
of the physics that shapes galaxy formation. It is easier to explore,
albeit qualitatively, somewhat more schematic solutions in SAMs, which
may point the way towards more physically rigorous investigations with
numerical techniques. 

SAMs have been used extensively to investigate and interpret
observations of nearby and distant galaxies in emission. A recent
generation of SAMs that incorporates feedback from accreting black
holes has been shown to be successful at reproducing a broad range of
observations. These include the stellar mass function and luminosity
function, gas fraction vs. stellar mass relation, and relative
fraction of early vs. late type galaxies as a function of stellar mass
at $z=0$, and the evolution of the global stellar mass density and
star formation rate density with redshift from $z\sim 6$ to 0
\citep{Bower2006, Croton2006, deLucia2007, Monaco2007, S08, Hopkins2009, Guo2010, Somerville2012}.
However, these models still fail to reproduce some important
observations. For example, they predict that low mass galaxies form
too early and are too quiescent at late times, reflecting star
formation histories that apparently do not match observational
constraints \citep{Fontanot2009, Weinmann2012}. Numerical
hydrodynamical simulations with similar implementation of ``sub-grid''
recipes largely show the same successes and problems
\citep{Weinmann2012}.  It has been suggested
\citep[e.g.][]{Fontanot2009,Krumholz2012} that inadequacies in our
modeling of star formation and/or stellar feedback are likely culprits
for these remaining difficulties in reproducing observations of
low-mass galaxies.

Meanwhile, recent observational and theoretical work has greatly
advanced our understanding of the physics that regulates star
formation on galactic scales. The vast majority of previous
cosmological simulations relied on the classical ``Kennicutt-Schmidt''
(KS) relation as a recipe for describing how cool gas turns into stars. The
KS relation, based on observations of nearby spiral galaxies and
starburst nuclei, says that the star formation rate surface density
($\Sigma_{SFR}$) is proportional to the total gas surface density
($\Sigma_{H} = \Sigma_{HI} + \Sigma_{H_{2}}$) \citep{Schmidt1959,
  Kennicutt1989, Kennicutt1998}. The KS relation is frequently
approximated as a power law, $\Sigma_{SFR} \propto \Sigma_{H}^N$, with
$N \sim 1.4$, above a critical total gas surface density $\Sigma_{\rm
  crit}$. Empirical studies have shown that $\Sigma_{\rm
  crit} \simeq 3$--10~ \msunpc \citep{Martin2001}.

However, \citet{Wong2002} showed that $\Sigma_{SFR}$ is more tightly
correlated with the density of \emph{molecular} hydrogen
$\Sigma_{H_{2}}$ (as traced by CO) than with the total gas
density. These results were confirmed and expanded upon with the
results from the THINGS survey \citep{Walter2008} combined with CO
maps from BIMA SONG and HERACLES \citep{Helfer2003, Leroy2009}. 
These studies showed that
$\Sigma_{SFR} \propto \Sigma_{H_{2}}^N$ with $N$ very close to unity,
implying that star formation takes place in molecular gas with roughly
constant efficiency \citep{Bigiel2008, Bigiel2011}. 
These results
underlined the importance of modeling the partitioning of gas into atomic
and molecular phases in theoretical models of galaxy formation.

\cite{Blitz2004, Blitz2006} showed that, empirically, the fraction of
molecular to molecular plus atomic gas, $f_{H_{2}}$, in nearby spirals
is tightly correlated with the disc midplane pressure.
\citet{Ostriker2010} proposed a theoretical explanation for this
relationship, arguing that the thermal pressure in the diffuse
Interstellar Medium (ISM), which is proportional to the UV heating
rate and therefore to the SFR, adjusts until it balances the midplane
hydrostatic pressure set by the vertical gravitational field. Other
recent theoretical work has argued that, as H$_2$ forms most
efficiently on dust grains, the metallicity of the gas, along with its
surface density, should be an important factor in determining
$f_{H_2}$ \citep{Krumholz2008, Krumholz2009_HI, Gnedin2010}.  Using
high resolution numerical simulations of isolated galaxies with
detailed chemistry and an H$_2$-based star formation recipe,
\citet{Robertson2008} showed that $f_{H_2}$ depended on metallicity,
gas surface density, and the UV background radiation.
\citet{Gnedin2010, Gnedin2011} characterized this dependence in detail
with high resolution numerical cosmological simulations with detailed
chemistry. The impact on the structural properties of disc galaxies of
using an H$_2$-based star formation recipe rather than a traditional
KS recipe has recently been explored with high resolution ``zoom-in''
cosmological simulations \citep{Christensen2012}.

Several SAM-based studies have modelled the partitioning of gas into
atomic and molecular phases using various approaches, and studied the
effect of using an H$_2$-based star formation recipe. \cite{Obreschkow2009}  
partitioned gas into atomic and molecular components using the
empirical pressure-based relation of \citet[][hereafter BR]{Blitz2006} in 
post-processing
on the Millennium semi-analytic models. \cite{Lagos2011} and \cite{Fu2010} 
implemented gas partitioning self-consistently into SAMs using two
approaches: the empirical pressure-based recipe of BR, and the
theoretically motivated metallicity-dependent recipe of 
\citet[][hereafter KMT09]{Krumholz2009}. 
These models then implemented an H$_2$-based star
formation recipe based on their computed H$_2$ fractions. Using a
similar approach, Somerville, Popping \& Trager (in prep; SPT14) explored the
partitioning of gas using the BR recipe, the KMT recipe, and an
additional metallicity dependent recipe provided by Gnedin \& Kravtsov
(GK), along with an H$_2$-based star formation recipe based on the
\citet{Bigiel2008} observational results. They concluded that the GK
recipe was more successful and robust, particularly at low
metallicities, than the KMT formulation \citep[see also][]{Krumholz2011}.
\citet[][hereafter PST14]{Popping2014} presented the predictions of
these models for the gas content of galaxies in \HI, H$_2$, and CO from
redshift six to zero for direct comparison with upcoming surveys of
gas tracers in emission.

The SAM developed by SPT14 does not predict the internal structure of
galaxies in detail, so we assume that the density profiles of the gas
and stellar discs are described by a smooth exponential function in
both the vertical and radial dimensions. We rely on simplified
approximations to estimate the scale length of the gas disc from the
specific angular momentum (spin) of the host dark matter halo. This
approach has been shown to reproduce the evolution of stellar disc
sizes (as traced by their optical light) from $z\sim 2$ to 0
\citep{Somerville2008a}, and also reproduces the observed sizes of \HI\
discs in the nearby universe, the observed sizes of CO discs in local
and high redshift galaxies for the small sample currently available,
and the spatial extent of the SFR density in nearby and high-redshift
galaxies (PST14). However, we also consider models in which the gas is
more extended than in our standard models, either because the gas that
forms the disc or is accreted onto the disc has higher specific
angular momentum than the dark matter halo \citep[as some numerical
  simulations suggest; e.g.][]{Robertson2004, Robertson2006, Agertz2011,
  Guedes2011}, accreted gas from cold streams deposit their angular 
  momentum to the inner parts of the halo \citep{Kimm2011},   or the gas 
  is in a non-rotationally supported extended configuration such as tidal 
  tails or an outflow \citep{Stewart2011b, Stewart2013}.

In this paper, we make use of the SAMs developed in SPT14 and PST14 to
explore for the first time the predictions for the properties of DLAs
in semi-analytic models with partitioning of gas into different phases
and an H$_2$-based star formation recipe. We investigate the impact of
the gas partitioning, star formation recipe, and assumptions about the
structure of the cold gaseous disc on the main observable properties
of DLAs and confront our predictions with the latest observations. The
paper is organized as follows. In section \ref{Sec:Sims}, we describe
the ``base-line'' semi-analytic models, the new recipes for
partitioning gas into an atomic and molecular component, and the new
H$_2$-based star formation recipes. We also describe our methodology
for generating gas distributions, and how we compile mock catalogs of
DLAs. In section \ref{Sec:Results}, we present our predictions for key
DLA observables, including the DLA column density distribution as a
function of redshift, DLA cross-section as a function of halo mass and
redshift, comoving density of DLAs and cosmological neutral gas
density as a function of redshift, distribution of DLA velocity
widths, DLA metallicity distribution, and DLA metallicity as a
function of velocity width and redshift. We discuss the implications
of our results in section \ref{Sec:Discussion}, and summarize and
conclude in section \ref{Sec:Conclusions}. Throughout this paper, we
adopt the following values for the cosmological parameters:
$\Omega_{\rm m}=0.28$, $\Omega_{\Lambda}=0.72$, $H_0=70.0$, $\sigma_8=0.81$,
and $n_s=0.96$. Our adopted baryon fraction is 0.1658.  These values
are consistent with the seven-year \textit{Wilkinson Microwave
  Anisotropy Probe} (\textit{WMAP}) results \citep{Komatsu2011}.  All
quoted metallicities are relative to solar.

\section{Models and Methodology}
\label{Sec:Sims}

\subsection{The Semi-Analytic Model of Galaxy Formation}

The semi-analytic models (SAMs) used here to compute the formation and
evolution of galaxies within a $\Lambda$CDM cosmology were originally
presented in \cite{Somerville1999} and \cite{Somerville2001}, with
significant changes described in detail in \citet[][hereafter S08]
{S08}, \citet[][hereafter S12]{Somerville2012}, and most
recently in SPT14.  The S12 SAM includes the following physically
motivated ingredients: (1) the growth of dark matter structure in a
hierarchical clustering framework as described by `merger trees', (2)
shock heating and radiative cooling of gas, (3) conversion of cold gas
into stars via an empirical `Kennicutt-Schmidt' relation, (4)
evolution of stellar populations, (5) a combination of feedback and
metal enrichment of the interstellar and intracluster medium from
supernovae, (6) `quasar' and `radio' mode black hole growth and
feedback from AGN, (7) starbursts and morphological transformation due
to galaxy mergers. Here, we briefly summarize these ingredients ---
for a more detailed description of the model framework, see S08 and
S12. The ingredients in the models used here are the same as
  those described in S12, with the exception of the new recipes for
  gas partitioning and star formation, which we describe below.
Throughout this work, we assume a standard $\Lambda$CDM universe
and a Chabrier stellar initial mass function
\citep[IMF;][]{Chabrier2003}.

The merging histories (or merger trees) of dark matter haloes are
constructed based on the Extended Press-Schechter formalism using the
method described in \citet{Somerville1999a}, with improvements
described in S08. These merger trees record the growth of dark matter
haloes via merging and accretion, with each ``branch'' representing a
merger of two or more haloes. We follow each branch back in time to a
minimum progenitor mass $M_{\rm res}$. We refer to $M_{\rm res}$ as
the mass resolution of our simulation where we have adopted $M_{\rm
  res} = 10^{9.5} \msun $ in all the models presented here. Our SAMs
give nearly identical results when run on the EPS merger trees or on
merger trees extracted from dissipationless N-body simulations 
\citep[][; Porter et al. in prep]{Lu2014}.

Whenever dark matter haloes merge, the central galaxy of the largest
progenitor becomes the new central galaxy, and all others become
`satellites'. Satellite galaxies lose angular momentum due to
dynamical friction as they orbit and may eventually merge with the
central galaxy. To estimate this merger timescale we use a variant of
the Chandrasekhar formula from \citet{boylan-kolchin:08}. Tidal
stripping and destruction of satellites are also included as described
in S08. We have checked that the resulting mass function and radial
distribution of satellites (sub-haloes) agrees with the results of
high-resolution N-body simulations that explicitly follow
sub-structure \citep{Maccio2010}.

Before reionization, each halo contains a mass of hot gas equal to the
univeral baryon fraction times the virial mass of the halo. After
reionization, which we assume to be complete by $z=11$, the
photoionizing background suppresses the collapse of gas into low-mass
haloes. We use the results of \cite{Gnedin2000} and \cite{Gnedin2004}
to model the fraction of baryons that can collapse as a function of
halo mass after reionization.

When a dark matter halo collapses or experiences a merger that more
than doubles the mass of the largest progenitor, the hot gas is
shock-heated to the virial temperature of the new halo. This gas then
cools and collapses based on a simple spherically symmetric model. We
assume that cold gas is accreted only by the central galaxy, even
though realistically, satellite galaxies should receive some fraction
of the new cold gas.

All newly cooling gas collapses to form a rotationally
supported disc. The scale radius is based on the initial angular
momentum of the gas and the halo profile, assuming angular
momentum is conserved and the self-gravity of the collapsing baryons
causes contraction in the inner part of the halo
\citep{Blumenthal1986, Flores1993, Mo1998}. \cite{Somerville2008a}
showed that this approach reproduced the observed size versus
stellar mass relation for discs from $z\sim 0$ to $2$. 
In this scenario, the cold gas specific angular momentum is set equal 
to that of the dark matter: $f_j = j_{\rm gas}/j_{DM} = 1$.

Star formation occurs in two modes, a ``normal'' mode in isolated
discs, and a merger-driven ``starburst'' mode. Star formation in the
``normal'' mode is modelled as described in
Section \ref{Sec:SFr} below. The efficiency and timescale of
the merger driven burst mode is a function of the merger mass ratio
and the gas fractions of the progenitors, and is based on the results
of hydrodynamic simulations of binary galaxy mergers
\citep{robertson:06a,hopkins:09a}.

Some of the energy from supernovae and massive stars is assumed to be
deposited in the ISM, resulting in the driving of a large-scale
outflow of cold gas from the galaxy. The mass outflow rate is
parameterized as a function of the galaxy circular velocity times the
star formation rate, as motivated by the ``energy driven'' wind scenario. 

Some fraction of this ejected gas
escapes from the potential of the dark matter halo, while some is
deposited in the hot gas reservoir within the halo, where it becomes
eligible to cool again. The fraction of gas that is ejected from the
disc but retained in the halo versus ejected from the disc and halo is
a function of the halo circular velocity (see S08 for details), such
that low-mass haloes lose a larger fraction of their gas.
The gas that is ejected from the halo is kept in a larger
``reservoir'', along with the gas that has been prevented from falling
in due to the photoionizing background. This gas is allowed to
``re-accrete'' onto the halo as described in S08.

Each generation of stars also produces heavy elements, and chemical
enrichment is modelled in a simple manner using the instantaneous
recycling approximation. For each parcel of new stars ${\rm d}m_*$, we
also create a mass of metals ${\rm d}M_Z = y \, {\rm d}m_*$, which we
assume to be instantaneously mixed with the cold gas in the disc. The
yield $y$ is assumed to be constant, and is treated as a free
parameter. When gas is removed from the disc by supernova driven winds
as described above, a corresponding proportion of metals is also
removed and deposited either in the hot gas or outside the halo,
following the same proportions as the ejected gas.

Mergers are assumed to remove angular momentum from the disc stars and
to build up a spheriod. The efficiency of disc destruction and
spheroid growth is a function of progenitor gas fraction and merger
mass ratio, and is parameterized based on hydrodynamic simulations of
disc-disc mergers \citep{hopkins:09a}. These simulations indicate that
more ``major'' (closer to equal mass ratio) and more gas-poor mergers
are more efficient at removing angular momentum, destroying discs, and
building spheroids. Note that the treatment of spheroid formation in
mergers used here has been updated relative to S08 as described in
\citet{Hopkins2009}. The updated model produces good agreement with
the observed fraction of disc vs. spheroid dominated galaxies as a
function of stellar mass \citep[][Porter et al. in prep]{Hopkins2009}.

In addition, mergers drive gas into galactic nuclei, fueling black
hole growth. Every galaxy is born with a small ``seed'' black hole
(typically $\sim 100\, \msun$ in our standard models). Following a
merger, any pre-existing black holes are assumed to merge fairly
quickly, and the resulting hole grows at its Eddington rate until the
energy being deposited into the ISM in the central region of the
galaxy is sufficient to significantly offset and eventually halt
accretion via a pressure-driven outflow. This results in
self-regulated accretion that leaves behind black holes that naturally
obey the observed correlation between BH mass and spheroid mass or
velocity dispersion \citep{dimatteo:05,Robertson2006,S08}.

There is a second mode of black hole growth, termed ``radio mode'',
that is thought to be associated with powerful jets observed at radio
frequencies. In contrast to the merger-triggered mode of BH growth
described above (sometimes called ``bright mode'' or ``quasar mode''),
in which the BH accretion is fueled by cold gas in the nucleus, here,
hot halo gas is assumed to be accreted according to the Bondi-Hoyle
approximation \citep{bondi:52}. This leads to accretion rates that are
typically only about $\la 10^{-3}$ times the Eddington rate, so that
most of the BH's mass is acquired during episodes of ``bright mode''
accretion. However, the radio jets are assumed to couple very
efficiently with the hot halo gas, and to provide a heating term that
can partially or completely offset cooling during the ``hot flow''
mode (we assume that the jets cannot couple efficiently to the cold,
dense gas in the infall-limited or cold flow regime).

\subsection{Multiphase Gas Partitioning}
\label{Sec:Gas_pre}

Throughout this paper we refer rather loosely to ``cold'' gas, which
is gas that according to our simple cooling model has been able to
cool below $10^4$ K via radiative atomic cooling. Most previous
cosmological simulations have considered all of this ``cold'' gas to
be eligible to form stars. Here, we partition it into components that
we label atomic, molecular, and ionized, and only allow the
``molecular'' component to participate in star formation.  As we do
not explicitly track the temperature or density of the ``cold'' gas in
our models, this is obviously still extremely schematic. However, when
we refer to ``cold'' gas, we are referring to gas that is in one of
these three states and is dynamically associated with the galactic
disc (rather than in an extended hot halo, an outflow, etc).

At each timestep, we compute the scale radius of the cold gas disc using
the angular momentum based approach described above, and assume that
the \emph{total} (\HI + \Htwo) cold gas distribution is described by an
exponential with scale radius $r_{\rm gas}$. 
We do not attempt to
track the scale radius of the stellar disc separately, but make the
simple assumption that $r_{\rm gas} = \chi_{\rm gas} r_{\rm star}$,
with $\chi_{\rm gas}=1.7$ fixed to match stellar scale lengths at
$z=0$. \citet{Bigiel2012} showed that this is a fairly good
representation, on average, for the discs of nearby spirals.
We then divide the gas disc into radial annuli and compute the
fraction of molecular gas, $f_{\rm H_2}(r) \equiv \Sigma_{\rm
  H_2}(r)/[\Sigma_{\rm H_2}(r)+\Sigma_{\rm HI}(r)]$, in each annulus,
as described below. 

\subsubsection{Ionized Gas}

Most (if not all) previous semi-analytic models have neglected the
ionized gas associated with galaxies, which may be ionized either by
an external background or by the radiation field from stars within the
galaxy. Here we include a simple analytic estimate of the ionized gas
fraction motivated by the model presented in \citet{Gnedin2012_BTF}. We
assume that some fraction of the total cold gas in the galaxy, $f_{\rm
  ion, int}$, is ionized by the galaxy's own stars. In addition, a
slab of gas on each side of the disc is ionized by the external
background radiation field. Gas is assumed to be ionized if it lies
below a critical threshold surface density $\Sigma_{\rm
  HII}$. Throughout this paper we assume $f_{\rm ion, int} = 0.2$ (as
in the Milky Way) and $\Sigma_{\rm HII} = 0.4 \, \msunpcsq$ (as in
Gnedin 2012). Applying this model within our SAM gives remarkably good
agreement with the ionized fractions as a function of circular
velocity shown in Fig.~2 of \citet{Gnedin2012_BTF}, obtained from
hydrodynamic simulations with time dependent and spatially variable 3D
radiative transfer of ionizing radiation from local sources and the
cosmic background.

\subsubsection{Molecular Gas: Pressure Based Partioning}

We consider two approaches for computing the molecular gas fractions
in galaxies. The first is based on the empirical pressure-based recipe
presented by \citet[][BR]{Blitz2006} who found that the molecular
fraction $R_{\rm mol} \equiv \Sigma_{\rm H2}/\Sigma_{\rm HI}$ is
correlated with the disc hydrostatic mid-plane pressure $P$:
\begin{equation}
R_{\rm mol} = \left(\frac{P}{P_0}\right)^{\alpha_{\rm BR}}
\end{equation}
where $P_0$ and $\alpha_{\rm BR}$ are free parameters that are
obtained from a fit to the observational data. 
We adopted $\log
P_0/k_B = 4.23$ cm$^3$ K and $\alpha_{\rm BR}=0.8$ from
\citet{Leroy2008}.

We estimate the hydrostatic pressure as a function of the distance
from the center of the disc $r$ as \citep{Elmegreen1989, Elmegreen1993, Fu2010}:
\begin{equation}
P(r) = \frac{\pi}{2} G \Sigma_{\rm gas}(r)[\Sigma_{\rm gas}(r) + 
f_{\sigma}(r) \Sigma_*(r)]
\end{equation}
where $G$ is the gravitational constant, $\Sigma_{\rm gas}$ is the
cold gas surface density, $\Sigma_*$ is the stellar surface density,
and $f_{\sigma}$ is the ratio of the vertical velocity dispersions of
the gas and stars:
\begin{equation}
f_{\sigma}(r) = \frac{\sigma_{\rm gas}}{\sigma_{*}} 
\end{equation}
Following \citet{Fu2010}, we adopt $f_{\sigma}(r) = 0.1
\sqrt{\Sigma_{*,0}/\Sigma_*}$, where $\Sigma_{*, 0} \equiv m_*/(2 \pi
r_*^2)$, based on empirical scalings for nearby disc galaxies.

\subsubsection{Molecular Gas: Metallicity Based Partioning}

\citet{Gnedin2011} performed high-resolution ``zoom-in'' cosmological
simulations with the Adaptive Refinement Tree (ART) code of
\citet{Kravstov1999}, including gravity, hydrodynamics,
non-equilibrium chemistry, and simplified radiative transfer. These
simulations are able to follow the formation of molecular hydrogen
through primordial channels and on dust grains, as well as
dissociation of molecular hydrogen and self- and dust-
shielding. These simulations also include an empirical \Htwo-based
star formation recipe.

\citet{Gnedin2011} presented a fitting function based on their
simulations, which effectively parameterizes the fraction of molecular
hydrogen as a function of the dust-to-gas ratio relative to the Milky
Way, $D_{\rm MW}$, the UV radiation field relative to the Milky Way
value $U_{\rm MW}$, and the neutral gas surface density
$\Sigma_{HI+H_2}$.  Following \citet{Gnedin2010}, we take the
dust-to-gas ratio to be equal to the metallicity of the cold gas in
solar units, $D_{\rm MW} = Z/Z_{\odot}$.  The UV background is
  defined as the ratio of the interstellar FUV flux at $1000{\mathrm
    \AA}$, relative to the Milky Way value of $10^6$ photons cm$^{-2}$
  s$^{-1}$ sr$^{-1}$ eV$^{-1}$ \citep{Draine1978, Mathis1983}, $U_{\rm
    MW}$.  In this work, we create two sets of models: one where the
  UV background is fixed to the Milky Way value ($U_{\rm MW}=1$)
  \citep{Murray2010, Robitaille2010} , and one where it is equal to
  the global star formation rate within the galaxy, $U_{\rm MW}=$ SFR
  M$_\odot$ yr$^{-1}$. 

The fitting functions from \cite{Gnedin2010} are intended to
characterize the formation of \Htwo\ on dust grains, which is the
dominant mechanism once the gas is enriched to more than a few tenths
of Solar metallicity. However, other channels for \Htwo\ formation in
primordial gas must be responsible for producing the molecular
hydrogen out of which the first stars formed. Studies with numerical
hydrodynamic simulations containing detailed chemical networks and
analytic calculations have shown that \Htwo\ can form through
primordial channels in dark matter haloes once they grow above a
critical mass of $M_{\rm III} \sim 10^5\, \msun$
\citep[e.g.][]{Nakamura2001, Glover2012_FS}. This gas can then
  form ``Pop III'' stars which pollute their surroundings and enrich
  the ISM to $Z_{\rm III} \sim 10^{-3} Z_\odot$ \citep{Schneider2002,
    Greif2010, Wise2012}.  Since these processes are thought to have
  taken place in haloes that are much smaller than our resolution
  limit, we represent them in a simple manner. We adopt a ``floor'' to
  the molecular hydrogen fraction in our haloes, $f_{\rm H2,
    floor}$. In addition, we ``pre-enrich'' the initial hot gas in
  haloes, and the gas that is accreted onto haloes due to cosmological
  infall, to a metallicity of $Z_{\rm pre-enrich}$. We adopt typical
  values of $f_{\rm H2, floor}=10^{-4}$ and $Z_{\rm pre-enrich} \sim
  10^{-3}\, \zsun$, based on the numerical simulation results
  mentioned above \citep{Haiman1996, Bromm2004}.  Our results are not
  sensitive to reasonable changes in these values, as shown in
  SPT14. 

\subsection{Star Formation}
\label{Sec:SFr}

The ``classical'' Kennicutt-Schmidt (KS) recipe \citep{Kennicutt1998}
assumes that the surface density of star formation in a galaxy is a
function of the \emph{total} surface density of the cold neutral gas
(atomic and molecular), above some threshold surface density
$\Sigma_{\rm crit}$. This approach has been used to model star
formation in most previous SAMs and numerical hydrodynamical
simulations. Here, we instead use a star formation recipe based on the
\Htwo\ content of the galaxy, motivated by recent observational
results.

\citet{Bigiel2008} found, based on
observations of spiral galaxies from the THINGS survey, that the
star formation timescale in molecular gas is approximately constant,
i.e.
\begin{equation}
\label{eqn:bigiel1}
 \Sigma_{\rm SFR} = A_{\rm SF} \, {\Sigma_{\rm H_2}}^{N_{\rm SF}}
\end{equation}
with $N_{\rm SF} \simeq 1$. 

Observations of higher density environments, such as starbursts and
high redshift galaxies, suggest that above a critical $H_2$ surface
density, the star formation timescale becomes a function of
$\Sigma_{\rm H_2}$ such that the star formation law steepens. Recent
work in which a variable conversion factor between CO and $H_2$ is
accounted for suggests that $N_{\rm SF} \simeq 2$ for high
$\Sigma_{\rm H_2}$ \citep{Narayanan2012b}. 
This steepening is also expected on theoretical grounds
\citep{Krumholz2009,Ostriker2011}. 
Therefore, in SPT14 we also considered an \Htwo-based
star formation recipe of the form

\begin{equation}
\label{eqn:bigiel2}
\Sigma_{\rm SFR} = A_{\rm SF} \, \left(\Sigma_{\rm H_2}/(10 \msun {\rm pc}^{-2}) \left(1+
\frac{\Sigma_{H_2}}{\Sigma_{\rm H_2, crit}}\right)^{N_{\rm SF}}\right)
\end{equation}

In SPT14, we found that the ``two-slope'' star formation recipe
produces better agreement with observations of star formation rates
and stellar masses in high redshift galaxies, so we adopt it in all of
the models presented in this work. For the parameter values,
  we adopt $A_{\rm SF}=6.0 \times 10^{-3}\, \msun ~{\rm yr}^{-1} ~{\rm
    kpc}^{-2}$, $\Sigma_{\rm H_2, crit} = 70 \msun$ pc$^{-2}$, and
  $N_{\rm SF}=1.0$, consistent with the observational results
  discussed above.

\subsection{Model Variants}
\label{Sec:modelvariants}

We consider nine main variants of our models: three recipes
  for gas partitioning (the pressure-based BR recipe and the
  metallicity-based GK recipe with a fixed/variable UV radiation
  field), and three choices for the specific angular momentum of the
  gas relative to the dark matter halo, parameterized by $f_j$. We
  consider fixed values of $f_j=1$ and $f_j=2.5$, and also a set of
  models in which $f_j$ is set based on the merger history of the
  galaxy. 
  \textbf{The $f_j=2.5$ models result in stellar and gaseous disks with higher specific angular momentum than their dark matter halos, and are motivated by numerical simulations that suggest this situation may arise due to stellar driven winds and/or cold flows (see Section \ref{Sec:Intro} and \ref{Sec:Disc_gas} for a more detailed discussion and references).} 
  In the merger models, we compute the disc properties and
  star formation rates using the $f_j=1$ models, then place the gas in
  a more extended distribution based on the halo's merger history, as
  we discuss further below.  These model variants are denoted
  GKfj1, GKj1, BRj1, GKfj25, GKj25, BRj25, GKfjm, GKjm, and BRjm and
  are summarized in Table \ref{Tab:models}.  While we only model
azimuthally symmetric extended cold gas discs, we consider them as a
proxy for other processes thay may cause the gas to be more extended.

Although we use the same \Htwo-based star formation recipe in all of
our models, both the choice of $f_j$ and the gas partitioning recipe
can affect the star formation efficiency. A larger value of $f_j$
leads to larger discs and lower gas densities overall, less efficient
formation of \Htwo\ and less efficient star formation. Similarly, the
different gas partitioning recipes lead to different \Htwo\ fractions
as a function of mass and redshift (see PST14) and therefore again to
higher or lower star formation efficiency, since only \Htwo\ can form
stars in our models.

\begin{table}
\centering
\caption{Model Definitions}
\begin{tabular}{l c c c c}
\hline \hline
Model   & f$_{H_2}$ & U$_{MW}$ & f$_j=j_{\rm gas}/j_{\rm DM}$ \\
\hline

GKfj1    & GK  & 1   & 1.0  \\
GKj1     & GK  & $\propto$SFR & 1.0  \\
BRj1     & BR  & - & 1.0  \\
GKfj25   & GK  & 1   & 2.5  \\
GKj25    & GK  & $\propto$SFR & 2.5  \\
BRj25    & BR  & - & 2.5  \\
GKfjm$^{*}$ & GK  & 1   & 1.0, 1.5, 2.5  \\
GKjm$^{*}$  & GK  & $\propto$SFR & 1.0, 1.5, 2.5  \\
BRjm$^{*}$  & BR  & - & 1.0, 1.5, 2.5  \\

\hline
\end{tabular}
\\
 $^* ~f_j= 1.0,~1.5,~2.5$ depending on if the galaxy has undergone no mergers, 
 only minor mergers, or at least one major merger respectively.
\label{Tab:models}
\end{table}

Our merger-based models are very crude and used to investigate the
impact of mergers on the angular momentum of the gas discs. In these
merger models, we begin with the BRj1 and GKj1 models respectively.
In post-processing, we boost the $f_j$ value to 1.5 or 2.5 after the
galaxy has had a minor or major merger, respectively. 
These models reflect the idea that some of the orbital angular
momentum of the merging galaxy may be transferred to internal (spin)
angular momentum following a merger.  This effect has been observed in
numerical simulations  \citep[e.g.][]{Robertson2006, Robertson2008,
  Sharma2012}, which suggest that major mergers have a larger effect
on the specific angular momentum distribution. Our $f_j$ values in
these models contain a small, but arbitrary offset, comparable to
their results.  One important inconsistency in the merger models is
that by increasing the cold gas angular momentum, we decrease the cold
gas density, which in turn will decrease the star formation
rate. Since $f_j$ is increased after running the semi-analytic model,
the stellar masses and star formation rates reflect those of the
$f_j=1$ models and are artificially high. 
For these reasons, we treat the `merger' models more as toy models
that provide some information on the different effects of the
distribution of cold gas on DLA properties; however, we focus the
majority of our analysis on the four other models.

The semi-analytic models contain a number of free
  parameters. These are kept fixed to the same values used in S12,
  except those involved in the star formation recipes, which are
  specified above. These parameter values were found to reproduce
  fundamental galaxy properties at $z=0$. As shown in SPT14 and
  \ref{Sec:MF_z0}, the updated star formation recipes have a very
  minor effect on the model results used for calibration, such as
  stellar mass function and gas mass functions.


\subsection{Selecting HI absorption systems}
\label{Sec:ModelDLA}

We obtain a catalog of host haloes by extracting haloes along lightcones
from the Bolshoi simulations \citep{Klypin2011, Behroozi2010}. These
lightcones cover a 1 by 1 deg$^2$ area on the sky over a redshift
range $0<z<5$ and contain galaxies with dark matter halo masses from
$10^{9.5}$ to $\sim 10^{14.5}$ M$_{\odot}$. 
However, the Bolshoi
simulation begins to become incomplete at $V_{vir} \simeq 50$ km
s$^{-1}$, $M_h \simeq 10^{10} \msun$; see
\cite{Klypin2011} for more details. These haloes are then populated
with galaxies as described above.

The molecular and atomic hydrogen gas is distributed in a disc with an
exponential radial and vertical profile. The vertical scale height is
proportional to the radial scale length, $z_{g} = \chi_z r_{g}$, where
$\chi_z = 0.4$ is a constant, in agreement with observations of
moderate redshift galaxies \citep{Bruce2012}. 
We explore different values of $\chi_z$, although reasonable values of 
$\chi_z$ (i.e. not razor-thin discs) have a minimal effect on our results.

\begin{figure*}
  \includegraphics[width=6in]{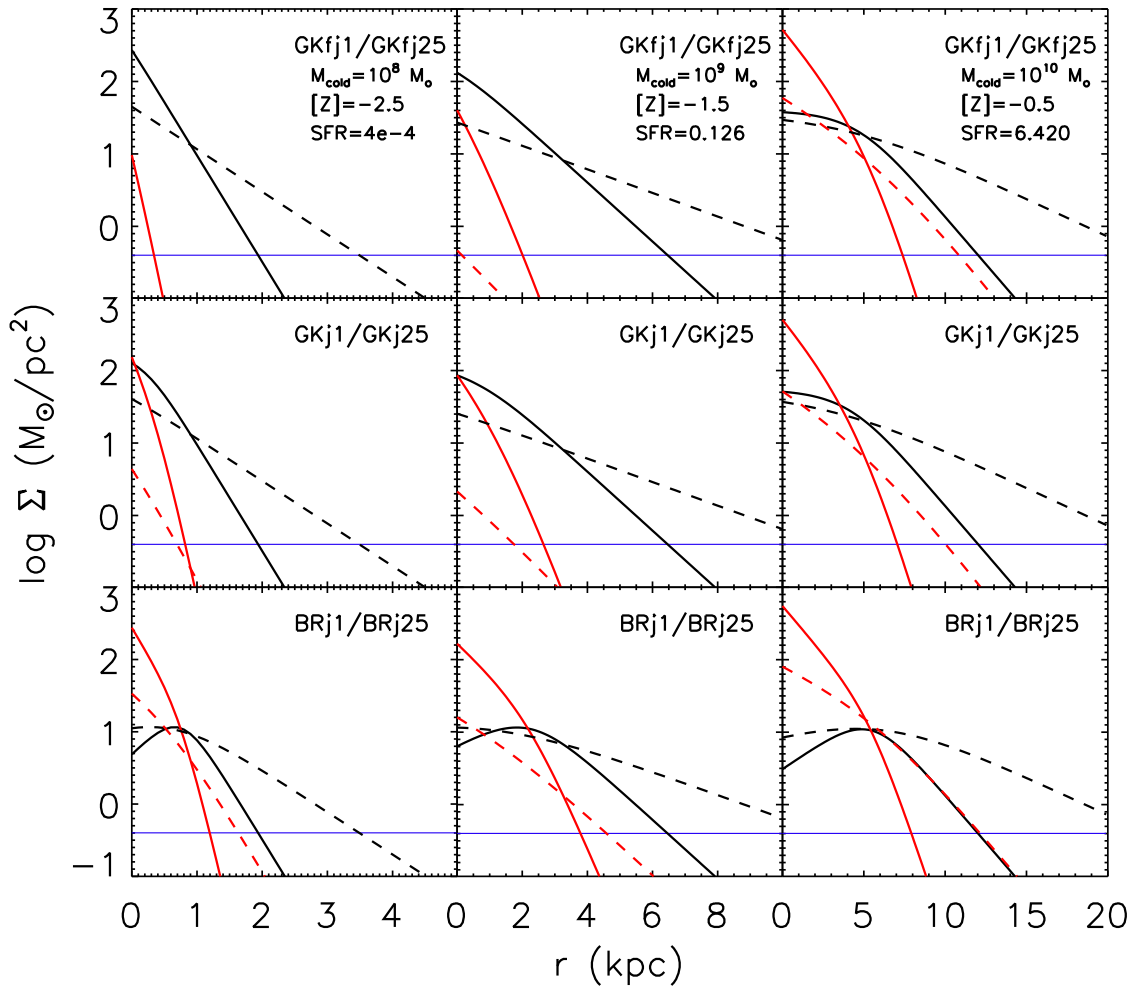}
\caption{ Cold gas radial profiles for three typical galaxies
    in the $f_j=1$ (solid line) and $f_j=2.5$ (dashed line) models
    with redshifts $2<z<3$. The HI gas (black), H$_2$ (red), and HII
    (blue) surface densities are shown. Cold gas mass and metallicity
    increase from left to right from ${\rm log} ~ Z = -2.5$ to $-0.5$
    and ${\rm log} ~ M_{\rm cold} / \msun = 8$ to 10. The top row
    shows the GKf models, the middle row the GK models, and the bottom
    row the BR models. Since stars form based on the density of H$_2$
    (red), this illustrates that the $f_j=1$ models are more efficient
    at forming stars than the $f_j=2.5$ models. Moreover, star
    formation is much more efficient in low mass halos in the BR and
    GK models than the GKf models (note the lack of any H$_2$ in the
    lowest mass galaxy in the GKfj25 model). However, the star
    formation efficiency becomes similar in all three models at high masses
    and metallicities. }
\label{Fig:Gas_prof} 
\end{figure*}

The central gas density is then defined as
$n_0 = M_{\rm cg} / (4 \pi \mu m_{H} r^{2}_{g} z_{g})$, where 
$M_{\rm cg}$ is the atomic and molecular gas, $m_{H}$ is 
the mass of the hydrogen atom, and $\mu$ is the mean molecular weight of the
gas. The atomic gas density as a function of radius along and 
height above the plane is given by
\begin{equation}
n_{\rm HI}(r,z) = n_0 \Bigl(1-f_{H_2}(r)\Bigr) \exp \Bigl(-\frac{r}{r_{g}} \Bigr) 
\exp \Bigl(-\frac{|z|}{z_{g}} \Bigr) 
\end{equation}

In Figure~\ref{Fig:Gas_prof} we show gas profiles for
three galaxies with cold gas masses of 
${\rm log} ~ M_{\rm cold}/ \msun = 8,~9,~10$ and metallicities 
${\rm log} ~ Z/ \zsun = -2.5,~ -1.5,~-0.5$ at $2<z<3$ 
for the f$_j$ = 1 and 2.5 models in the SAMs (as seen face-on). 
Each row shows the difference in cold gas partitioning for our three 
models and three fiducial galaxies. Star formation is much more efficient 
in low mass halos in the BR and GK models than the GKf models due to the 
high cold gas density threshold for H$_2$ formation in the latter. However 
once a significant amount of metals have been produced, the star formation 
efficiency converges in all three models as can be seen in galaxies with 
high masses and metallicities.
For reference, a neutral hydrogen column density of $N_{HI} = 2\times
10^{20}$ cm$^{-2}$ corresponds to a gas surface density of
$\Sigma_{HI} = 10$ M$_{\odot}$ pc$^{-2}$. 
Figure \ref{Fig:Gas_prof} demonstrates the impact on the gas
distribution of the different assumptions for gas partitioning, star
formation, and gas angular momentum.

The models provide the radial distance from the central galaxy for
each satellite galaxy, and we assign a random azimuth and polar angle
$\phi$ and $\theta$ for each satellite's position with respect to the
central.  With the positions determined for every galaxy in each
lightcone, we generate 20,000 random sightlines and integrate the
three-dimensional gas density distribution along the sightline.  Each
galaxy is given a random orientation angle with respect to the
sightline.  All sightlines as well as the properties of all haloes with
observed column densities above a threshold of $\NHI > 10^{19}~
\atomscm $ are then saved as our catalog of absorption systems.

We then generate low-ionization line profiles by assuming the gas is
distributed in small clouds within the disc, using a similar approach
to that of \cite{Maller2001}. The relevant parameters are:
$\sigma_{\rm int}$, the internal velocity dispersion of each cloud; $N_c$,
the number of clouds; and $\sigma_{\rm cc}$, their isotropic random
motions. Following PW97, we take $\sigma_{\rm int}=4.3$ km s$^{-1}$.  PW97
derived this value from Voigt profile fits to their observations with
$N_c=5$ being the minimum acceptable number of components. Increasing
the number of clouds to as high as 60 did not improve the goodness of
fit for a disc model since the model discs are relatively
thin. Following \cite{Maller2001}, we assume the gas discs are cold
and set $\sigma_{\rm cc}=10$ km s$^{-1}$. We combine these internal
velocities with the rotational velocity of the disc as well as the
relative orbital velocity of the satellite galaxy with respect to the
central (when applicable). For each sightline, we treat the gas
density distribution as a continuous probability distribution for the
positions of the clouds \citep{Maller2001}. We then generate low
ionization line profiles by randomly distributing 20 clouds with the
same optical depth along the line of sight. Finally, we `measure' the
velocity width, $\Delta v_{90}$, by taking the difference between the
pixel containing 5\% and 95\% of the total optical depth.

In generating the low-ionization line profiles, we make a number of
simplifying assumptions. Satellite galaxies are assumed to be on
circular orbits. Gas discs are assumed to have a simple radial profile
in addition to being axisymmetric. The gas distribution is independent
of galaxy environment or Hubble type. We do not account for distortion
in gas discs due to the gravitational effects of other galaxies or
effects due to previous merger events, except very crudely in the
merger (``m'') models as described above.

\section{Results}
\label{Sec:Results}

In this section, we compare the predictions for our suite of models
with a set of observations of DLAs.
To calibrate our models, in section \ref{Sec:MF_z0} we present the
$z=0$ stellar, HI, and \Htwo\ mass functions for our models along with
observations from local galaxies.  In section \ref{Sec:DLA_prop} -
\ref{Sec:cs_Mh}, we present column density distribution functions, the
comoving line density of DLAs, the cosmological neutral gas density in
DLAs ($\Omega_{DLA}$), and DLA cross sections and halo masses as a
function of redshift for all of our models and high-redshift DLAs. The
DLA metallicity distribution, the cosmic evolution of DLA
metallicities, the effects of metallicity gradients, and DLA
kinematics are presented in sections \ref{Sec:Z_dv90} and
\ref{Sec:dv90}.  In sections \ref{Sec:Z_dv90} and \ref{Sec:dv90}, we
only consider the $f_j=1$ and 2.5 models as the merger-based models
have very similar metallicities to the $f_j=1$ models. Additionally,
we feel that our merger-based models are too crude to meaningfully
attempt to predict the kinematics.  We focus the majority of
  our analysis on the GKj25 and BRj25 models, since the $f_j=1$ models
  fail to reproduce the column density distribution, the number of
  DLAs, and the mass of HI in DLAS (although the $f_j=1$ models are
  actually the closest to the `fiducial' model presented in previous
  SAMs, e.g. S08 and S12). The GKfj25 model produces a large number of
  low mass ``pristine'' halos, which experience no star formation and
  so contain gas close to the pre-enriched metallicity, which we
  believe to be unphysical.  Note that in this paper, we focus on the
observational properties of the DLAs themselves. In a follow-up paper,
we will present the optical properties of the DLA host galaxies in our
models (Berry et al. 2014, in prep.).

\subsection{Local Stellar and Cold Gas Mass Functions}
\label{Sec:MF_z0}

The usual approach used in semi-analytic models is to calibrate the
models using a subset of observations of $z \sim 0$ galaxies. The
galaxy stellar mass function and cold gas fractions, or mass functions
of cold gas, are commonly used quantities for this calibration
procedure. A more extensive comparison with observations for the
GKfj1, GKj1, and BRj1 models is presented in SPT14 and
PST14. In addition, the BRj1 model produces extremely similar
predictions to the models presented in S08 and S12. Here we examine
the impact of varying the H$_2$ formation recipe and the distribution
of cold gas ($f_j$) on several fundamental calibration quantities: the
local galaxy stellar mass function (GSMF), HI mass function (HIMF),
and H$_2$ mass function (H2MF).

\begin{figure*}
  \includegraphics[width=6in]{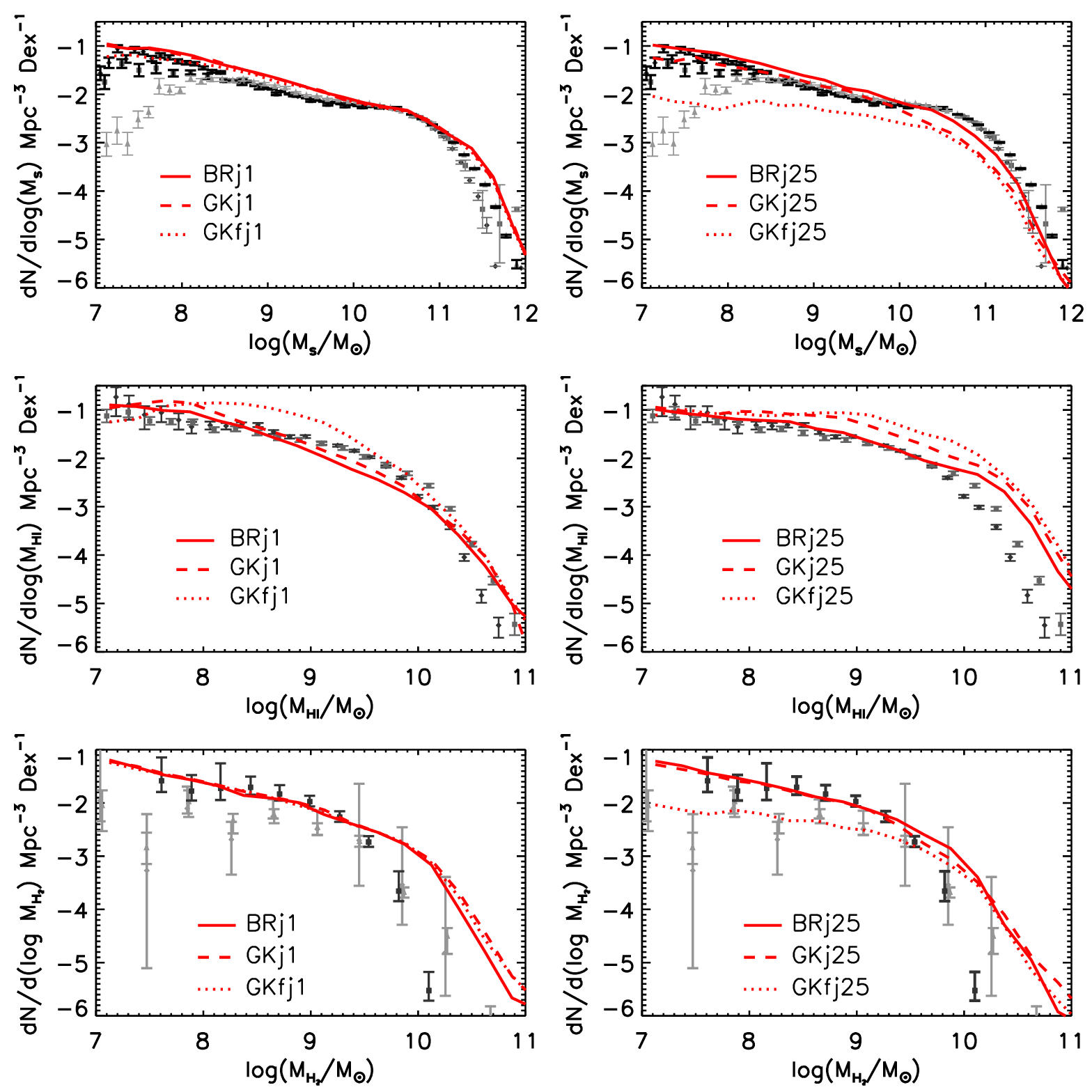}
\caption{Top left - galaxy stellar mass functions (GSMFs) for all
  galaxies in the GKfj1 (dotted lines), GKj1 (dashed lines) 
  and BRj1 (solid lines) models at $z=0$. 
  The local GSMFs are overplotted, with \protect\citet{Baldry2008,
    Baldry2012} in bold above and \protect\citet{Panter2007} and
  \citet{Li2009} in bold below log M$_\ast / \msun = 10.65$ to emphasize
  where each is more reliable. Middle left - same as top left except for 
  the \HI\ mass functions at $z=0$, with the local HIMF from
  \protect\citet[][gray]{Zwaan2005} and \protect\citet[][black]{Martin2010}
  overplotted. Bottom left - same as top left except for 
  the   H$_2$ mass functions (H2MFs) at 
 $z=0$, with the local H2MFs for a constant and variable $X_{CO}$
  factor from \protect\citet[][gray]{Keres2003} and
  \protect\citet[][black]{Obreschkow2009} overplotted. Right column - similar
  to left column except for the GKfj25 (dotted lines), GKj25 (dashed lines) 
  and BRj25 (solid lines) models. 
  The middle panels highlight how the \HI\ gas content of
  galaxies can help to constrain galaxy formation models. }
\label{Fig:MF_z0}
\end{figure*}

Figure \ref{Fig:MF_z0} shows the $z=0$ GSMFs, HIMFs, and H2MFs
  for all of our models. We do not show the GKfjm, GKjm, and BRjm mass
  functions as they are the same as the GKfj1, GKj1, and BRj1 models
  respectively, since the gas is redistributed only in
  post-processing.  As can be seen in the top row, the predicted $z=0$
  stellar mass function is extremely similar for the $f_j=1$ models,
  and is in reasonable agreement with observations of the local GSMF.
  The $f_j=2.5$ models tend to produce too few galaxies with large
  stellar masses, with the largest discrepancy around the knee of the
  GSMF. Of the $f_j=2.5$ models, the BRj25 model is in the best
  agreement with the observed GSMF.  In the models with extended gas
  distributions, the lower gas densities cause star formation to be
  less efficient. In the GK models, H$_2$ formation is more efficient
  in gas with higher metallicity, but more H$_2$ is photo-dissociated
  if the UV background is high. In the GK model with a fixed UV
  background, star formation becomes very inefficient in low mass,
  low-metallicity halos. In the GK model with a varying UV background,
  the UV radiation field intensity is also lower in these low-mass
  halos, which goes in the opposite direction, leading to a larger net
  fraction of H$_2$ and therefore less suppression of star formation
  relative to the GKf model.

We also find that the GK and BR extended disk ($f_j=2.5$)
  models produce reasonable agreement with the observed GSMF and
  galaxy star formation rate function for galaxies selected via their
  stellar emission at $z\sim 2$. We show these results along with a
  more detailed comparison of our model predictions with the optical
  properties of DLA host galaxies in Berry et al. (2014, in prep.). 

The middle row of Figure \ref{Fig:MF_z0} shows the HIMFs for the
models along with local 21-cm observations from the HIPASS and ALFALFA
surveys \citep{Zwaan2005, Martin2010}, which highlights the power of
using cold gas observations to discriminate between models.  None of
our models matches the observed \HI\ mass function well in detail.
The BRj1 and GKj1 models provide the best match to the
  observations, but produce slightly too few systems with intermediate
  \HI\ masses ($10^8 ~\msun \lesssim M_{\rm HI} \lesssim 10^{10}
  ~\msun$). The GKfj1 model overproduces the number of systems with
  $M_{HI} \lesssim 10^{9.5}~ \msun$. The BRj25 and GKj25 models are
  more successful at reproducing the slope of the observed HIMF at low
  masses, but significantly overpredict the number of systems with
  high HI masses.  The GKfj25 model produces too many galaxies at all
  HI masses.  In general, the model HIMFs show that galaxies in the
  fixed-UV GK models have more HI than in the varying-UV and BR
  models, for the reasons discussed above.  Similarly, galaxies with
more extended gas distributions have more HI and shallower slopes for
their HIMFs than the traditional disc models.

The bottom row of Figure \ref{Fig:MF_z0} shows the $z \sim 0$ H2MFs
for the models along with the inferred H2MF from the FCRAO 
Extragalactic CO survey 
\citep{Keres2003} assuming a constant $X_{CO} = 2 \times 10^{20}$ factor 
and using a variable $X_{CO}$ factor as computed by
\citet{Obreschkow2009}.  The H2MFs for the $f_j=1$ models are
almost identical to each other and are in good agreement with both
observational estimates at low M$_{H2}$, but overproduce the number of
high-M$_{H2}$ systems. 
The predictions of the BRj25 and GKj25 models 
are very similar to the $f_j=1$ models, but have a slightly better 
fit at high-H$_2$ mass. the GKfj25 model produces substantially fewer
systems with low M$_{H2}$, leading to a flatter H2MF low-mass end
slope.
For all four models, the high mass end of the H2MFs are in better
agreement with the observational estimates of \cite{Keres2003}, which
assumed a constant conversion factor between CO and \Htwo
($X_{CO}$). However, the estimates obtained by \cite{Obreschkow2009}
with a variable $X_{CO}$ are likely to be more accurate. In
\cite{Keres2003}, the highest mass bin contains a number of CO
luminous starburst galaxies. PST14 provide a more detailed comparison
between the observed CO luminosity function and the $f_j=1$ models.

In addition, PST14 show a comparison of the radial sizes of
  galaxies in the $f_j=1$ models with observations, finding good
  agreement for the HI radii and SFR half-light radii from $z=0$ to
  2. The $f_j=2.5$ models produce SFR half-light radii that are still
  consistent with observations at $z=0$, but are about a factor of two
  larger than the $f_j=1$ model disks at $z=2$, in apparent conflict
  with observations. However, it is unknown to what extent the
  observed $z\sim2$ star forming galaxies may be biased towards
  compact objects, due to selection. PST14 also show the ratio of HI
  mass to stellar mass, ratio of \Htwo\ mass to stellar mass, and
  ratio of HI to \Htwo mass, as a function of galaxy stellar mass and
  surface density, showing good agreement with observations for $z=0$
  disk-dominated galaxies in the $f_j=1$ models. We have carried out
  this comparison for the $f_j=2.5$ models as well, and find that
  relative to the $f_j=1$ models they tend to produce slightly higher
  gas fractions overall, and slightly less \Htwo\ relative to
  HI. However, the results are still within the uncertainty on the
  observational values. 

\subsection{Column Density Distribution}
\label{Sec:DLA_prop}

The \HI\ column density distribution function, $f(\NHI , X)$, is one of
the best constrained observational quantities for \HI\ absorption
systems. It is defined as the number of absorbers with column
densities in the range $[\NHI, \NHI\ + d\NHI]$ per comoving
absorption length $[X, X + dX]$

\begin{equation}
f_{HI}(\NHI, X) d\NHI\ dX = n_{DLA}(\NHI, X)
\end{equation}

where $dX = \frac{H_0}{H(z)}(1+z)^2 dz$. Absorption systems with a 
constant comoving density and proper size have a constant density per 
unit $X$ along the sight-line. Observations indicate only mild evolution 
in the column density distribution function with redshift 
\citep[e.g.][]{Prochaska2009}.

\begin{figure}
  \includegraphics[width=3.4in]{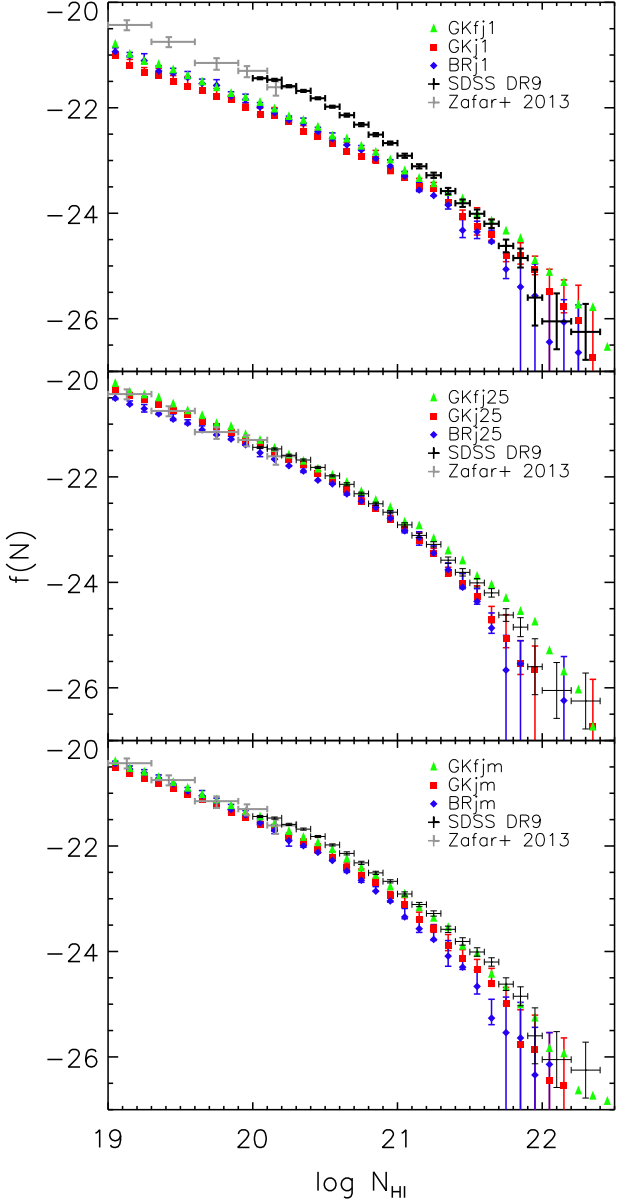}
\caption{Top panel - column density distribution function for
    \HI\ absorption systems in the $f_j=1$ models (GKfj1: green
    triangles; GKj1: red squares; BRj1: blue diamonds) in the redshift
    range $2<z<3.5$. Middle panel - same as top except for the $f_j=2.5$
    models. Bottom panel - same as top except for the merger
    models (GKfjm: green triangles; GKjm: red squares; BRjm: blue
    diamonds).  The model errors represent the maximum field-to-field
    variation for four subfields of $0.5 \times 0.5$ deg$^{-2}$. The
    fixed- and varying-UV GK models have similar errors. The SDSS DR9
  results from \protect\citet{Noterdaeme2012} are overplotted in black
  and the sub-DLA results from \protect\citet{Zafar2013b} in gray. The
  $f_j = 1$ models have a shallower slope in f(\NHI), causing them to
  produce too few low-\NHI\ systems, which is most apparent at ${\rm
    log} ~\NHI\ <21$. The $f_j = 2.5$ models provide the closest fit
  and are a good match to the data.  }
\label{Fig:fN_N_j1}
\end{figure}

The top panel of Figure \ref{Fig:fN_N_j1} shows the predicted column density
distribution function at $2<z<3.5$ for the $f_j=1$ models in the
range $10^{19}<\NHI\ <10^{22.5}$ cm$^{-2}$ compared with the recent SDSS 
data release (DR9) 
results from \cite{Noterdaeme2012} and observations of sub-DLAs
($10^{19}<\NHI$) from \cite{Zafar2013b}. 
We can see that all models do
moderately well at reproducing the number of high column density
systems, but greatly underproduce the lower column density
systems. This result has been shown before by \citet{Maller2001}, and
may indicate that the gas discs in the $f_j=1$ models are too
compact. An alternative is that there are a large number of DLAs that
are hosted in haloes below our resolution limit or that do not arise
from gas in galactic discs, although neither of these effects seems to
be very likely to make a large contribution, based on recent results
from numerical simulations \citep[e.g.][]{Fumagalli2011, Cen2012}. In
addition, we find this scenario to be unlikely as they would have
small velocity widths, inconsistent with observations.

Historically, no DLAs were known with column densities
log N$_{HI}>10^{22}$ cm$^{-2}$, and simulations had difficulty
reproducing this very sharp cutoff \citep[e.g.][]
{Nagamine2004_abund,Pontzen2008}. Recently, larger volume surveys such
as SDSS DR9 have revealed that although rare, these high HI column
density systems do exist. We note that, in the paradigm in which
metallicity is a fundamental parameter controlling the \HI-H$_2$
transition, it is more likely to produce high \HI\ column density
systems at high redshift as the threshold density for forming H$_2$
becomes higher for lower metallicity gas \citep[e.g.][]{Schaye2004,
  Erkal2012}. In our models, this is reflected in the larger numbers
of high column density systems predicted in the metallicity-dependent GK
models. We include \HII\ gas, but find it makes no significant
difference to the predicted column density distribution.

Motivated by the discrepancies in the number of low column density
systems in the models with $f_j=1$, we explore a simple model with a
more extended distribution of cold gas with $f_j=2.5$.  The middle
panel in Figure \ref{Fig:fN_N_j1} shows the column density
distribution for the GKfj25, GKj25, and BRj25 models.  These
`extended gas' models do significantly better than the $f_j=1$ models
at matching the observed column density distribution function,
reproducing the general shape of the column density distribution over
a wide range of column densities.  
The BRj25
model is not as successful at reproducing the number of DLAs at all
column densities especially at high-\NHI\, specifically at log
$\NHI\ \gtrsim 21.6$, although uncertainties due to cosmic variance
are larger in this regime.  Again, all models produce DLAs with log
$\NHI\ \gtrsim 22$ cm$^{-2}$, although they are rarer in the BRj25
model than the GK models, due to the differing amount of gas and the
density threshold for the \HI-\Htwo\ transition, discussed further
below.  The success of the $f_j=2.5$ models suggests that either the
gas that forms discs has higher specific angular momentum than the
dark matter halo material, or DLAs arise from gas in an alternate
extended distribution such as an outflow or tidal tails, although we
have not specifically modeled these configurations here. The picture
of DLAs arising from extended gas is supported by numerical
simulations, which have shown that stellar driven winds can
preferentially remove low-angular momentum material, leading to a
higher average specific angular momentum
\citep[e.g.][]{Brooks2011}. In addition, the gas specific angular
momentum can also be boosted by cold flows and mergers
\citep{Robertson2006, Agertz2011, Stewart2013}.

To explore the possible boosting of specific angular momentum by
mergers, we also consider a simple merger-based model in which $f_j$
is increased following major and minor mergers, as described in
Section~\ref{Sec:modelvariants}. The resulting column density
  distribution for the GKfjm, GKjm, and BRjm models is shown in the
  bottom panel of Figure \ref{Fig:fN_N_j1}. Interestingly, these
simple models do fairly well at reproducing the column density
distribution over the whole range shown, much better than the $f_j=1$
models, although they slightly underproduce the number of DLAs at all
$\NHI$. As they contain the same amount of \HI\ as the $f_j=1$ models,
their success suggests that the cold gas may be in an extended
distribution in a subset of galaxies due to the conditions of their
formation.

At the low-\NHI\ end of the column density distribution, sub-DLAs 
($19 < {\rm log} ~ \NHI\ < 20.3$) in the
$f_j=2.5$ models are in agreement with the results of \cite{Zafar2013b}, 
although our results become more uncertain at ${\rm log} ~ \NHI\ \lesssim 19.5$.  
Low column density systems are more likely to have been produced in
outflows and filaments of cold gas. Furthermore, the distribution of
gas in exponential discs likely does not extend smoothly down to
arbitrarily low-$\NHI$, and haloes below our mass resolution (log
M$_h/~ \msun < 9.5$) may also make a significant contribution to
sub-DLAs.  Therefore, we restrict the rest of our analysis to systems
selected as DLAs ($\log ~\NHI\ > 20.3$ \atomscm) as the majority are
likely produced in cold dense gas that is closely associated with
galaxies.

\begin{figure}
  \includegraphics[width=3.4in]{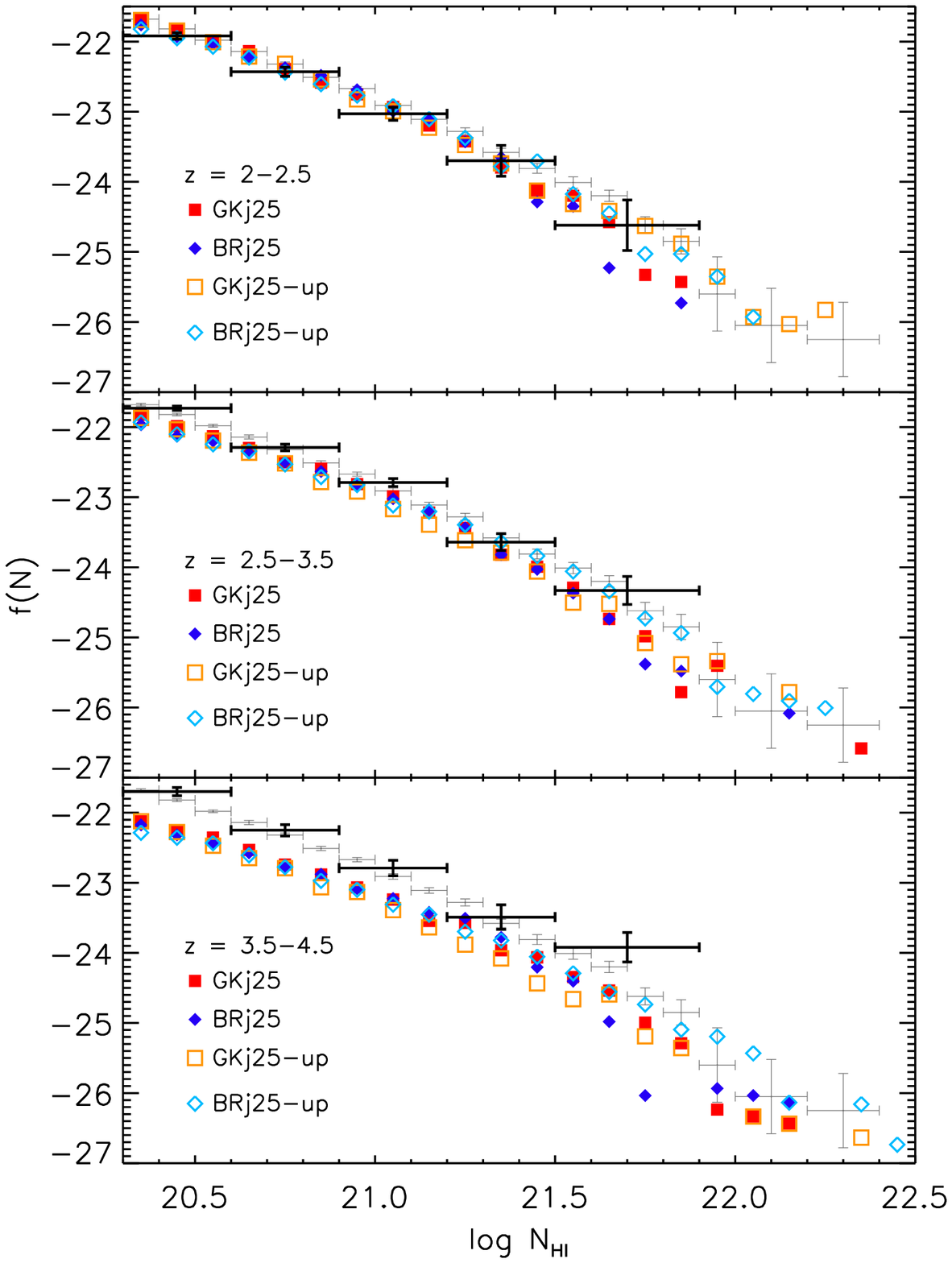}
  \caption{Similar to Figure \ref{Fig:fN_N_j1} except the column
    density distribution function is plotted for DLAs at redshifts
    $2<z<2.5$ (top), $2.5<z<3.5$ (middle), and $3.5<z<4.5$ (bottom) in
    the GKj25 (red squares) and BRj25 (blue diamonds)
      models. Open points show the total cold neutral gas column
      density including H$_2$. Observations from SDSS DR5
    \protect\citep{Prochaska2009} at the same redshifts are
    overplotted (black), along with the SDSS DR9 results at $2<z<3.5$
    (gray) for reference. Both the models and observations show a
    flattening of the column density distribution at higher
    redshifts. Each model shows a decline in the number of
    low-\NHI\ DLAs with redshift yet a comparable number of
    high-\NHI\ DLAs. At $3.5<z<4.5$, observations produce this
    flattening with more high-\NHI\ DLAs. Additionally, the
    unpartitioned models indicate that the HI-H$_2$ transition becomes
    important at log $\NHI \sim 21.7$, in qualitative agreement with
    the observations.}
\label{Fig:fN_N_z}
\end{figure}

Figure \ref{Fig:fN_N_z} shows the column density distribution function
at $2<z<2.5$, $2.5<z<3.5$, and $3.5<z<4.5$ for the GKj25 and BRj25
models with the SDSS DR5 observations at the same redshifts
overplotted \citep{Prochaska2009}.  The models at $2<z<3.5$ are
consistent with observations, although at $3.5<z<4.5$, both produce
fewer DLAs than are observed. The shape of the column density
distribution functions become flatter at higher redshifts in both
models, consistent with observations. However, in the models this
flattening results in a reduced number of low-\NHI\ systems in the
highest redshift bin, which is in conflict with observations. 
All of our models fail to reproduce the observed number of DLAs at $z
\gtrsim 3$ (see section \ref{Sec:Om_z}), and we see this here
reflected in the column density distribution.

\begin{figure*}
  \includegraphics[width=6in]{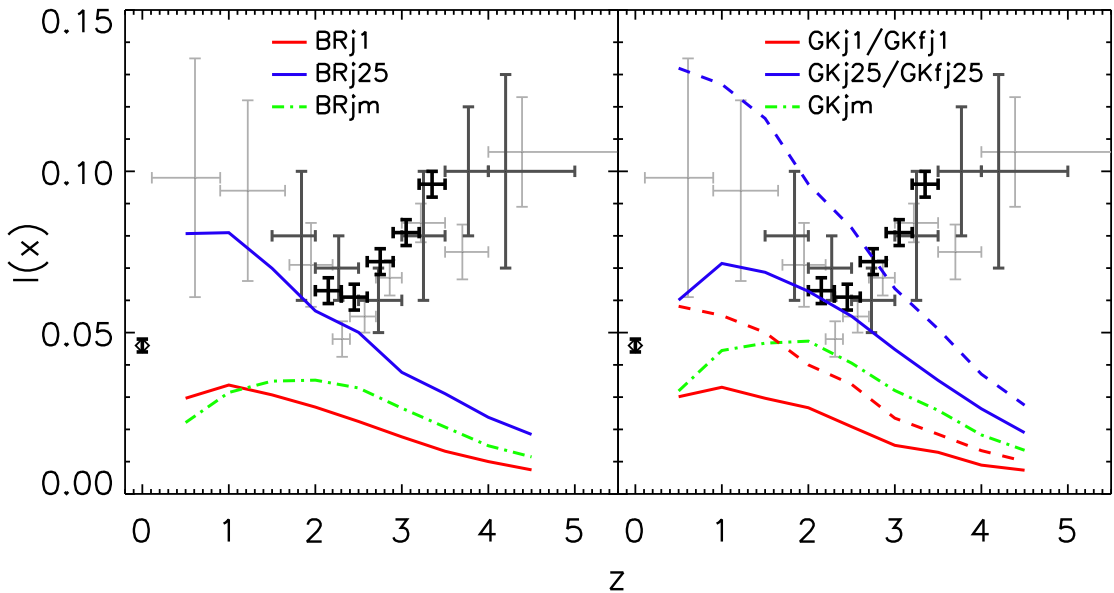}
    \caption{Left panel - the comoving line density of DLAs as
        a function of redshift for the BRj1 (red), BRj25 (blue), and
        BRjm (green dot-dashed) models. Right panel - same as the left
        except for the GKj1, GKjm, and GKj25 models with the GKfj1 and
        GKfj25 models (dashed) overplotted. Observations of Mg II
      absorbers \protect\citep{Rao2006} and high-redshift DLAs from
      \protect\citet{Prochaska2009} are overplotted in light gray
      while those from \citet{Zafar2013b} are shown in dark gray.
      Observations of local galaxies \protect\citep{Zwaan2005b} and
      high-redshift DLAs \protect\citep{Noterdaeme2012} are
      overplotted in black. All models fail at $z > 3$, and the
      $f_j=1$ models produce too few DLAs at all redshifts. The BRj25
      and GKj25 models are in the best agreement with the observations
      at $z\lesssim 2.5$, and the merger models are only a modest
      improvement over the $f_j=1$ models. }
    \label{Fig:lx_z}
\end{figure*}

We also show the column density distribution function of the 
BRj25 and GKj25 models where gas is left unpartitioned, BRj25-up and 
GKj25-up respectively. In these models, the total cold neutral gas density,
regardless of whether it is in HI or \Htwo, is used to compute the
column density, as in most previous models. These models allow us to study 
the effect of multiphase partitioning on the column density 
distribution function. 
We can see by comparing the partitioned and unpartitioned
 models that the \HI -\Htwo\ transition does lead to a
slightly steeper drop in the number of high column density
systems. This transition can be seen in Figure \ref{Fig:fN_N_z} at log
$\NHI \sim 21.7$, qualitatively consistent with observations. The small 
number of high-\NHI\ systems causes there to be significant scatter at 
high column densities. 
Both models produce more DLAs with very high column densities at 
higher redshifts, while these very high-\NHI\ DLAs are only seen in the 
unpartitioned models at $z \sim 2$, suggesting that the HI-H$_2$ transition 
may occur at lower density at high redshift.

\subsection{Comoving line density and $\Omega_{\rm g}({\rm z})$}
\label{Sec:Om_z}

The column density distribution function gives the number of DLAs per 
unit absorption path length for a given column density. The zeroth moment 
of this distribution is the line density of DLAs, which measures the 
number of DLAs per comoving absorption distance:

\begin{equation}
l_{\rm DLA} (X) dX = \int^{\infty}_{N_{20.3}}~f_{\rm HI}(N,X) dN~dX
\end{equation}

Figure \ref{Fig:lx_z} shows the comoving line density of DLAs as a
function of redshift for the BR models (left) and fixed- and varying-UV 
GK models (right). Observational estimates of
the line density of high redshift DLAs from \cite{Prochaska2009} and
\cite{Noterdaeme2012}, and that inferred from Mg II absorbers from
\cite{Rao2006} are also overplotted. As compared to the $f_j=1$
models, the larger \HI\ masses in the BRj25 and GKj25 models, discussed
in Section \ref{Sec:MF_z0}, are also reflected in the larger number of
DLAs at all redshifts, in much better agreement with observations at 
$z \lesssim 2.5$. The compact gas distributions of galaxies in 
the $f_j=1$ models cannot reproduce the observed number of DLAs
at any redshift. 
Additionally, the merger-based models are only a modest improvement 
over the $f_j=1$ models. 
Relative to the other models, the GKfj25 model gives
rise to significantly more DLAs at all redshifts.  As we will see
later, a large number of DLAs in the GKfj25 model are hosted in low
mass dark matter haloes. These systems have low metallicity, and in the 
GKf models, they are inefficient at converting gas into \Htwo\ and
subsequently into stars, so they have large \HI\ masses. Therefore a
large number are selected as DLAs, boosting the line density. 
The BRj25 and GKj25 models produce the best agreement with the 
data at $z<3$.

At $z>3$, all of our models produce far too few DLAs, and show the
opposite trend as observations (the number density of DLAs decreases,
rather than increases, with increasing redshift).  The reasons for
this fairly dramatic discrepancy are unclear. Two possible
explanations are that an increasing number of DLAs are associated with
gas in filaments or outflows at higher redshifts, or that the
distribution of gas in galactic discs evolves over cosmic time. Note
that the gas would have to be \emph{more extended} at higher redshifts
to alleviate this discrepancy.

\begin{figure*}
  \includegraphics[width=6in]{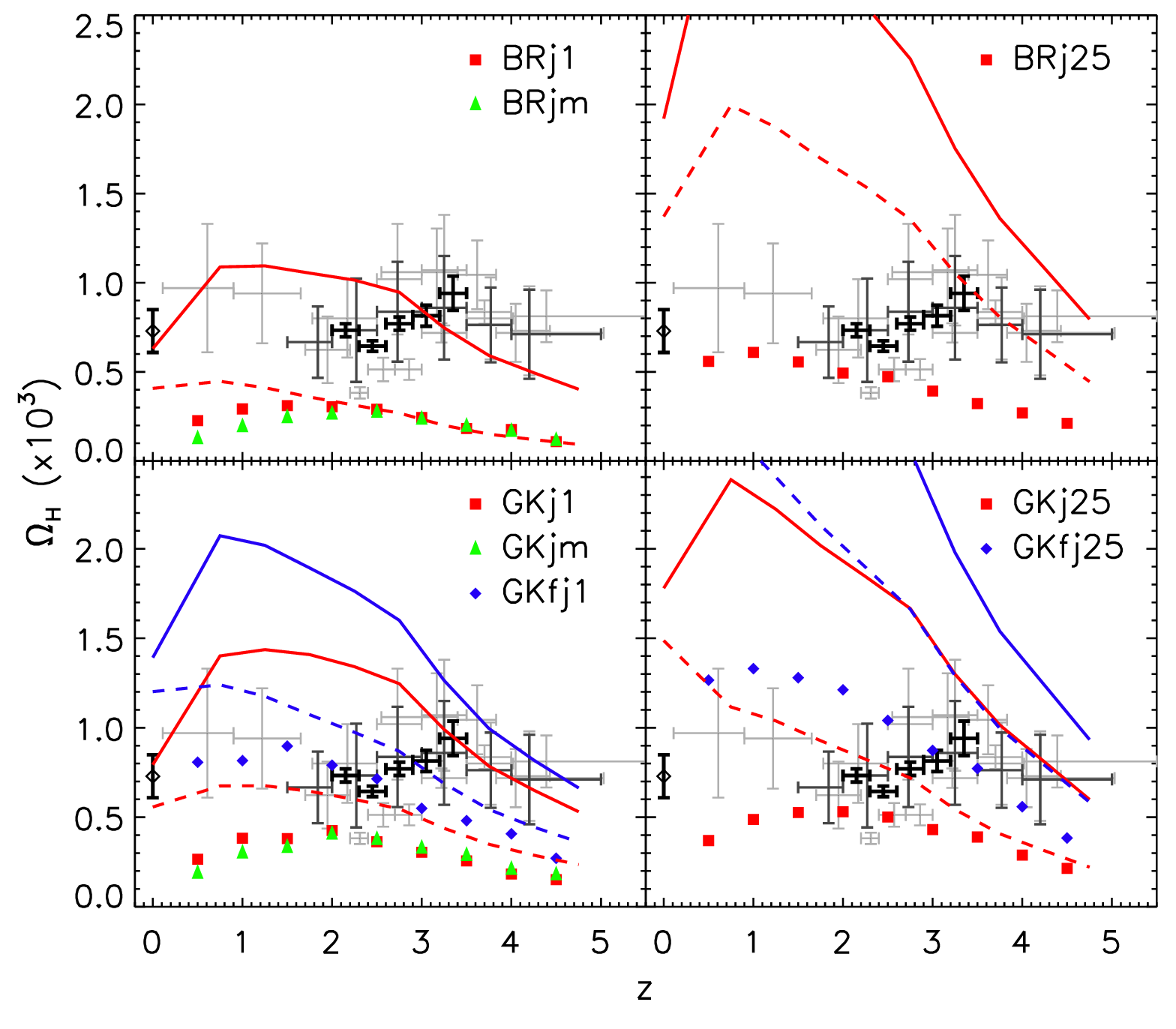}
\caption{The cosmic density of \HI\ contained in DLAs ($\Omega_{\rm
    DLA}$) as a function of redshift (solid shapes) and the amount of
  \HI\ in all galaxies ($\Omega_{\rm HI}$; dashed line) along with the
  total amount of cold neutral gas in all galaxies $\Omega_{HI+H_2}$
  (solid line). Observations of DLAs and Mg II absorbers
  \citep{Zwaan2005, Peroux2005, Rao2006, Prochaska2009,
    Noterdaeme2009, Guimaraes2009} are overplotted in light gray. The
  dark gray observations of DLAs and sub-DLAs are from
  \citet{Zafar2013b}.  Both the SDSS DR9 observations of high-redshift
  DLAs \citep{Noterdaeme2012} and observations of local galaxies
  \citep{Braun2012} are shown in black. Upper left panel - BRj1 (red
  squares) and BRjm (green triangles) models. Upper right panel -
  BRj25 model. Lower left panel - GKfj1 (blue diamonds), GKj1
    (red squares), and GKjm (green triangles) models. Lower right
    panel - GKfj25 (blue diamonds) and GKj25 (red squares) models. The
    BRj25, GKj25, and GKfj1 models are in reasonable agreement with
    observations at $z \lesssim 2.5$ while the GKj1, BRj1, and
    merger-based models underproduce $\Omega_{\rm DLA}$ at all
    redshifts.  The GKfj25 model is the only model that is remotely
    close to the data at $z\gtrsim 2.5$, but it predicts $\Omega_{\rm
      DLA}$ somewhat higher than observational measurements at lower
    redshift, and the column density distribution is in conflict with
    observations.  }
\label{Fig:Om_z}
\end{figure*}

\cite{Fumagalli2011} and \cite{Cen2012} found that large amounts of
DLA column density gas arise in filaments extending up to the virial
radius at $z = 4$. This fraction of DLA column density gas in
filamentary structures is significant at $z \gtrsim 3$ and decreases
monotonically with cosmic time, in keeping with the discrepancy
between our models and observations.  Moreover, the majority of
missing DLAs in our models are at low-$\NHI$, as would be expected for
intergalactic DLAs. If intergalactic DLAs, produced for example in
filaments of cold gas, make up a significant fraction of the DLA
population, then many DLAs will not be included in our models. 
Alternatively, if a significant number of high-redshift
DLAs are associated with haloes of mass log (M$_h/ ~ \msun ) < 10$, then
the discrepancy might be a resolution effect since our simulations are 
incomplete below this mass. Since DLA metallicities
at these redshifts are very low, and the formation of H$_2$ in this
regime is not well-understood, the amount of neutral hydrogen gas in a
given halo and the number of DLAs may be affected.


Using the column density distribution function and the comoving line
density of DLAs, we can calculate the total neutral hydrogen gas mass
density in DLAs using:

\begin{equation}
\Omega_{\rm DLA} = \frac{m_H H_0}{c \rho_{c,0}} \frac{\Sigma N(HI)}{\Delta X}
\end{equation}

where $m_H$ is the mass of the hydrogen atom, $H_0$ is the Hubble
constant, $\rho_{c,0}$ is the critical density at $z=0$, and the sum
is calculated for systems with log N(HI)$> 20.3$ across a total
absorption pathlength of $\Delta X$. Figure \ref{Fig:Om_z} shows the
total cold gas density ($\Omega_{\rm HI+H_2} (z)$), neutral
hydrogen density in all galaxies ($\Omega_{\rm HI} (z)$), and the neutral
hydrogen density inferred from systems selected as DLAs
($\Omega_{DLA}$) in the BRj1 and BRjm (top left) models; 
BRj25 (top right) model; GKfj1, GKj1, and GKjm (bottom left) models; 
and GKfj25 and GKj25 (bottom right) models.
Observational estimates of $\Omega_{\rm DLA}(z)$ from DLAs and Mg II
absorbers are overplotted for reference \citep{Peroux2005, Rao2006,
  Noterdaeme2009, Prochaska2009, Guimaraes2009, Braun2012, Noterdaeme2012}.  
As can be seen in Figure \ref{Fig:Om_z}, the 
$f_j=1$ and merger-based models underproduce the amount of 
$\Omega_{\rm DLA}(z)$. The GKfj1, 
GKj25 and BRj25 models produce the best fit to the $\Omega_{\rm DLA}(z)$
distribution. The GKfj1 model produces too few DLAs and 
too many high-\NHI\ DLAs (see Figure \ref{Fig:fN_N_j1}), causing it to 
be a coincidence that it reproduces 
the observed amount of $\Omega_{DLA}$. On the other hand, the GKfj25 model
produces too much \HI\ while the BRj1, GKj1, and merger-based models produce too
little \HI\ at these redshifts. This is consistent with the
conclusions drawn from their respective column density distribution
functions and comoving line densities. 
As already anticipated from Figure \ref{Fig:lx_z}, all of
our models contain less \HI\ in DLAs than is observed at $z \gtrsim
3$. Only the GKfj25 model is marginally consistent with the
observations at these redshifts.

Note that the different models make different predictions for the
total amount of cold gas in galaxies, as well as for the amount of
\HI\ in galaxies and the fraction of \HI\ in systems that would have
been selected as DLAs. The GKfj25 model predicts the largest amount of
cold gas overall, as well as the highest values of $\Omega_{\rm HI}
(z)$ and $\Omega_{\rm DLA}(z)$. This is because in this model, a lot
of gas has low metallicity and is at low surface density, leading to
inefficient \Htwo\ formation and star formation in many systems. It is
interesting to note that while the total cold gas density and
$\Omega_{\rm HI}$ tend to rise with decreasing redshift in all models, the
fraction of gas in systems that are selected as DLAs decreases,
leading to a flatter dependence of $\Omega_{\rm DLA}$ on redshift, in
better agreement with observations.  Overall, the $f_j=2.5$ 
models predict a lower fraction of HI to be contained in DLAs than the
$f_j=1$ models.  As high-$\NHI$ systems make the largest
contribution to $\Omega_{\rm DLA}$ and the $f_j=1$ models have relatively
more high-\NHI\ DLAs due to a flatter column density distribution
function, we expect a larger fraction of the total cold gas to come
from the central regions of galaxies.  Although there are more DLAs in
the GKjm and BRjm models, the reduced number of high-$\NHI$ systems is
evident as $\Omega_{\rm DLA}(z)$ for the GKjm model is comparable to the
GKj1 and BRj1 models at all redshifts.  An overproduction of
high-$\NHI$ DLAs in the BRj1 model causes the inferred amount of HI in
DLAs at $z\sim 4$ to be slightly \emph{larger} than the total amount
in all galaxies.  In spite of a significant decrease in the number
density of DLAs at $z>3$ in all models, the $\Omega_{\rm DLA}(z)$
distribution remains relatively flat. This result is in agreement with
the flattening of the column density distribution as was discussed in
section \ref{Sec:DLA_prop}.

Returning to the discrepancy between our model predictions and
observations at $z \gtrsim 3$, it is first interesting to note that in
the BRj1 model, even the \emph{total} cold gas density at $z \gtrsim
3$ is lower than the observational estimates of $\Omega_{\rm HI}$ from
DLAs. Indeed, this model is quite similar to the model presented in
S08, and this discrepancy has already been pointed out in that work
(see their Figure 14). It can also be seen from the results presented
in S08 that the predicted $\Omega_{\rm cold}$ at high redshift is
quite sensitive to the assumed cosmological parameters, in particular
the power spectrum normalization $\sigma_8$. This suggests that part
of the problem may be due to too-efficient star formation and/or
overly efficient ejection of gas by strong stellar winds at these
epochs in these models. 

At redshifts $z<2$, all of the models predict a relatively flat
dependence of $\Omega_{\rm DLA}$ on redshift, in qualitative agreement
with observations, although the normalization is too low in the
BRj1/BRjm models and a bit high in the GKfj25 model. This is the case
even in models (BRj25, GKj25) with much more rapidly rising total gas
density and $\Omega_{\rm HI}$. The large amount of \HI\ in galaxies that
would not be selected as DLAs in the BRj25 and GKj25 models arises
from \HI\ in lower column density systems in low mass haloes (log M$_h~/~
\msun < 10$), which have low gas surface densities and small DLA cross
sections.  

Taken together, our model results suggest that the rather flat
dependence of $\Omega_{\rm DLA}$ on cosmic time from $4.5 \lesssim z
\lesssim 1$ derived from observations of DLAs could be a cosmic
coincidence: at $z\gtrsim 3$, $\Omega_{DLA}$ may be `contaminated' by
cold gas that is not closely associated with galaxies, while at lower
redshifts ($z\lesssim 3$), $\Omega_{DLA}$ may significantly
underestimate the total atomic gas content of all galaxies. 
These results show the danger in assuming that 
$\Omega_{DLA} = \Omega_{\rm cold}$ or even $\Omega_{\mathrm HI}$. 

\subsection{DLA Halo Masses and Cross-sections}
\label{Sec:cs_Mh}

\begin{figure*}
  \includegraphics[width=6in]{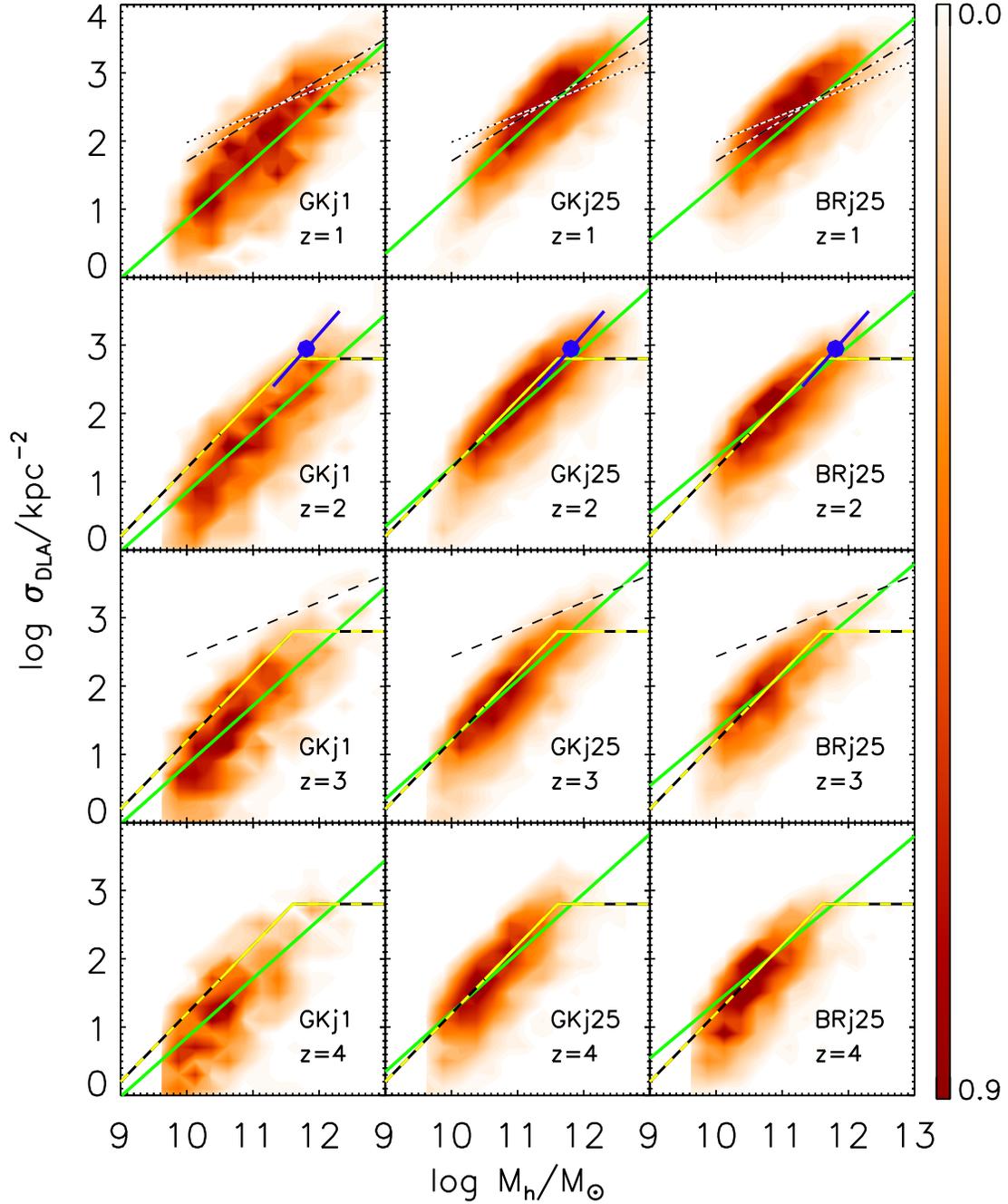}
    \caption{Number density distribution of DLA cross section versus
      halo mass for our sample of DLAs in the GKj1 (left column),
      GKj25 (middle column), and BRj25 (right column) models at
      $0.5<z<1.5$ (top row), $1.5<z<2.5$ (second row), $2.5<z<3.5$
      (third row), $3.5<z<4.5$ (bottom row) with the best-fit slope of
      the $\sDLA -M_h$ relation for each model at $z=2$ (green
      line). At high halo masses, the $\sDLA -M_h$ relation
      flattens. The data point (blue circle and line) from
      \citet{Font-Ribera2012} shows their estimate for $\sDLA$ at
      $<z>~=2.3$ with the best-fit $\sDLA -M_h$ slope. We also show
      the range of \sDLA\ values and halo masses at $z \sim 3$ for
      several sets of numerical simulations: the best-fit power law at
      $1.4< z < 4$ of \citet[measured, yellow; extrapolated,
        yellow-black]{Fumagalli2011}; and the best-fit power laws at
      $z=1$ (dotted), $z=1.6$ (dot-dashed), and $z=3$ (dashed) from
      \citet{Cen2012}. Compared to \citet{Font-Ribera2012},
        \sDLA , M$_h$, and $\alpha$ in the GKj25 model are in the best
        agreement.  Conversely, the $f_j=1$ models produce \sDLA\ and
        $\alpha$ values that are significantly lower than
        \citet{Font-Ribera2012}. }
\label{Fig:cs_Mh}
\end{figure*}

The DLA cross section represents the area in kpc$^2$ for which a
galaxy's gas surface density (corrected for inclination) would be high
enough for it to be selected as a DLA. It is straightforward to
compute this quantity in our models, as using our assumed density
profile we can easily compute the projected area within which the
column density is greater than ${\rm log} ~\NHI\ > 20.3$, which is its
DLA cross section. 
Figure \ref{Fig:cs_Mh} shows the distribution of DLA cross sections 
as a function of halo mass 
for our sample of DLAs in the GKj1, GKj25, and BRj25 models at
redshifts $z=1$, $z=2$, $z=3$, and $z=4$. We only show these models as 
each of the $f_j=1$ models has a similar distribution of DLA cross sections 
at a given halo mass as the GKj1 model. 
The GKj25 and GKfj25 models are also similar. 

We can see that in all
models and at all redshifts, DLAs are predicted to occupy haloes with a
fairly broad range of masses, $10^{10} \msun \lesssim M_h \lesssim
10^{12} \msun$. Moreover, the DLA cross-section versus halo mass
relation evolves mildly with time in any of the models. 
This has implications for DLA kinematics which we explore in a later
section.

The DLA cross-section at a given halo mass grows with cosmic time in
all of our models. By $z = 1$, DLAs in the $f_j=2.5$ models have halo
masses and DLA cross sections that are both typically $\sim 1.5$
decades larger than at higher redshift while they both span a similar
dynamic range.  Conversely in the $f_j=1$ models, there is a
significant fraction of small, compact DLAs at all redshifts and
evolution is seen as an increase in the number of higher mass DLAs.
Additionally, our merger tree mass resolution limit of $M_{\rm res} =
10^{9.5}\, \msun $ and the completeness limit of the host halo catalog
($V_{\rm vir} = 50$ km s$^{-1}$) also significantly reduces the number of
low mass DLAs.  These effects are relatively small at low redshift,
especially in the $f_j=2.5$ models. However the average halo mass
decreases with increasing redshift causing the mass resolution of our
models to become increasingly important, especially in the $f_j=1$ models.

\begin{table*}
\caption{Halo mass vs. DLA cross-section ($M_h - \sigma_{DLA}$) Relation}
\begin{tabular}{l c c c c c c c}
\hline \hline
 & GKfj1 & GKfj25 & GKj1 & GKj25 & BRj1 & BRj25 & FR12 \\
 \hline
$<\mathrm{(log ~M_h / M_\odot)}>$ & $4.9 \times 10^{10}$  & $5.3 \times 10^{10}$  & $8.3 \times 10^{10}$  & $1.1 \times 10^{11}$  & $8.5 \times 10^{10}$  & $8.6 \times 10^{10}$  & $6 \times 10^{11}$ \\
$< \sDLA / \mathrm{kpc^{-2}} >^{\ast}$ & 490$\pm$450 & 1030$\pm$740 & 570$\pm$660 & 1120$\pm$850 & 570$\pm$660 & 900$\pm$710  & 1400  \\
$\alpha$                            & 0.86 & 0.86 & 0.90 & 0.91 & 0.78 & 0.88 & 1.1$\pm0.1$ \\
\hline
\end{tabular}
\\
Our results for DLAs at $\langle z\rangle=2.3$ correspond to the same mean redshift as \cite{Font-Ribera2012}. Note, $\alpha$ changes with halo mass where higher mass halos have flatter $M_h - \sigma_{DLA}$ relations. \\
$^{\ast}$ for DLAs with $11.8 < {\rm log~ M_h}/\msun < 12.2$ where the errors show the scatter about the mean. 
\label{Tab:cs_Mh}
\end{table*}

Recently using observations of DLAs at $2<z<3.5$ from the BOSS survey, 
\citet[][hereafter FR12]{Font-Ribera2012} 
found that DLAs with a mean redshift of $<z>=2.3$, have a large range
of halo masses with an average halo mass of M$_h = 6 \times 10^{11}~
\msun$. For DLAs residing in haloes of mass M$_{h} \sim 10^{12}~
\msun$, they also find a mean DLA cross-section of $\sDLA\ = 1400$
kpc$^{2}$, and a $\sDLA-$M$_h$ relation that scales as $\sigma_{DLA}
\propto$ M$_h^\alpha$ where $\alpha=1.1 \pm 0.1$ with a minimum halo
mass of M$_h = 10^9~ \msun$.  In order to make an accurate comparison
to \cite{Font-Ribera2012}, we select all DLAs in each of our models
with redshifts $2<z<2.6$, which corresponds to a mean redshift of $\langle 
z\rangle=2.3$.  Table \ref{Tab:cs_Mh} shows our M$_h$, $\sDLA$, and $\alpha$
values for DLAs in each of our models. Our $f_j=2.5$ models produce
DLAs with halo masses and DLA cross sections that are the most similar
to \cite{Font-Ribera2012}. 
The BRj25 and GKj25 models produce slopes of
$\alpha \sim 0.9$, significantly flatter than that calculated in FR12. 
The second row of Figure
\ref{Fig:cs_Mh} shows model DLAs that are at a comparable redshift to
these observations.  This also shows that the $f_j=2.5$ models
produce the most comparable DLA cross sections in massive haloes $M_h \simeq
10^{12}\, \msun$ as FR12. The $f_j=1$ models produce significantly lower 
values of $\sDLA$. 

\cite{Font-Ribera2012} also find 
that DLA halo mass does not correlate with column density. This
result indicates that the column density distribution function has a
similar shape at low and high halo mass.  When we divide our sample in
half based on halo mass (${\rm log}~ M_h/\msun > 10^{11}$, ${\rm log}
~ M_h/\msun < 10^{11}$), we also find no correlation between DLA halo
mass and column density. The results of \cite{Font-Ribera2012} strongly 
support the picture of a significant population of DLAs at $z \sim 2.3$ 
arising from extended gas associated with more massive galaxies. 

We also compare our results to predictions from several different
numerical hydrodynamic simulations. We overplot the results from
\cite{Fumagalli2011} at $z=1.4-4$, and at
$z=1.0,~ 1.6, ~3.1$ from \cite{Cen2012} in Figure \ref{Fig:cs_Mh}. 
DLAs in the GKj1 model are more compact than DLAs observed in any
numerical simulation at any redshift.  In contrast, the BRj25 and
GKj25 models are in very good agreement with the results of
\cite{Fumagalli2011}. Our $f_j=2.5$ models are
in fair agreement with the predictions of \cite{Cen2012} at $z=1$ while 
DLAs in our models have a much steeper $\sDLA-$M$_h$ relation. 
At $z\sim 3$, \cite{Cen2012} finds much larger $\sDLA$ values at a 
given halo mass than our
models or the other simulations predict. This appears to be due to 
outflows boosting the DLA cross-section. Our models do not directly 
model outflows, although they may indirectly contribute to our 
extended disk models. Note
that an increasing contribution to the DLA cross-section from outflows
or filaments with increasing redshift could manifest in just this way,
as larger values of $\sigma_{\rm DLA}$ at a given halo mass.

At $z\sim 4$, more DLAs may arise in haloes below our resolution limit
since our models show a decrease in halo mass and $\sDLA$ with
redshift, especially in the GKj1 model. 
However, the gas fraction in haloes with $\log M_h/\msun \lesssim 9.5$ 
also drops rapidly due to the ``squelching'' of gas infall by the 
photoionizing background after re-ionization implemented in our models. 
At $z<3$, the steep slope 
and low fraction of low-halo mass DLAs (M$_h \sim 10^{10} ~\msun$) in the
$f_j=2.5$ models suggests that there are likely not many DLAs arising
in haloes below our resolution limit. 
If $\sDLA$ at a given halo mass also increased with
redshift, then we would expect even more DLAs to arise in these low
mass haloes. Both \cite{Fumagalli2011} and \cite{Cen2012} discuss the
contribution of DLAs arising from streams and clumps to the DLA
population at higher redshifts, although the simulations from 
\cite{Cen2012} only probe haloes more massive than M$_h > 10^{10.5}~ \msun$.  
At $z \sim 3-4$, both find that a large fraction of DLAs originate in
filamentary structures and gas clumps extending as far as the virial
radius. Conversely at $z<3$, these intergalactic DLAs make a much
smaller contribution to the total DLA population, in keeping with the
picture suggested by our results.

\subsection{Metallicities}
\label{Sec:Z_dv90}

\begin{figure}
\includegraphics[width=3.4in]{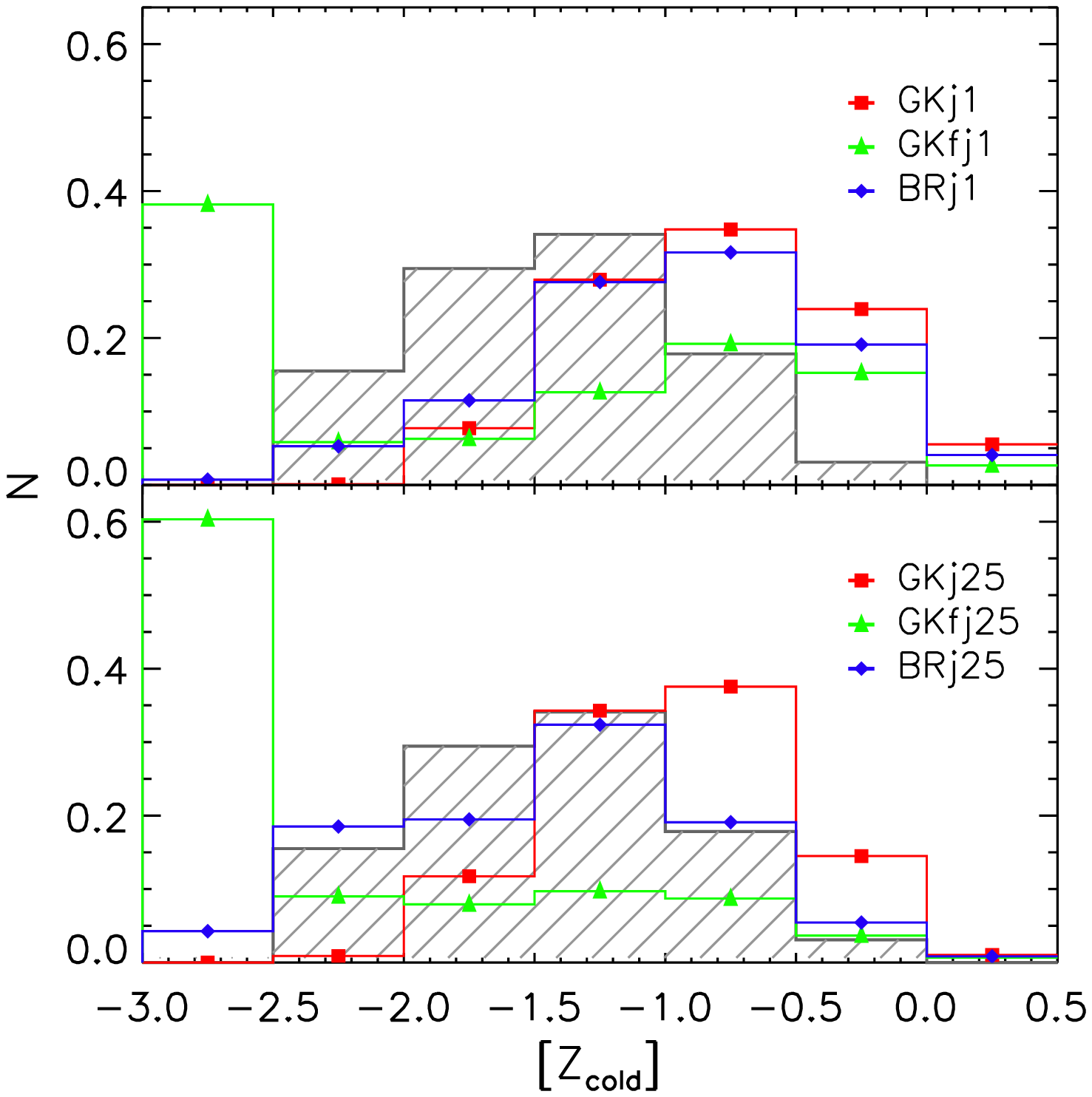}
\caption{Upper panel - distribution of metallicities for all
    DLAs in the redshift range $2<z<3.5$ for the GKfj1 (red), GKj1
    (green), and BRj1 (blue) models in the upper panel. Lower panel -
    same as top except for the GKfj25 (red), GKj25 (green), and BRj25
    (blue) models.  The observed distribution of DLA metallicities in
    the same redshift range are overplotted in gray
    \citep{Rafelski2012}. All the models except GKfj25 produce a
    distribution with a width similar to that of the observed one,
    with the BRj25 model producing the best agreement with the
    observed DLA metallicity distribution. The GKfj1 and GKfj25 models
    produce a significant number of DLAs with metallicities near the
    ``pre-enriched'' metallicity (${\rm log} ~Z = -3$).}
\label{Fig:Z_Mh}
\end{figure}

Figure \ref{Fig:Z_Mh} shows the distribution of cold gas-phase
metallicities for all galaxies identified as DLAs at $2<z<3.5$ 
in all of our models compared with SDSS-DR3 and SDSS-DR5 results from 
observed DLAs in the same redshift range from \citet{Rafelski2012}.  In this initial 
set of plots, we show the mass-weighted mean metallicity of the cold gas in
our model galaxies. Later, we consider the effects of metallicity
gradients.

Our $f_j=1$ and GKj25 models
show a roughly lognormal distribution of DLA metallicities with a peak
around log Z $\sim$ -0.8.  Both the fixed-UV models show flatter distributions 
with a large fraction of DLAs with metallicities near the
pre-enriched metallicity of log Z $ \sim -3$, indicating that a
substantial number have never undergone significant star formation. We
discuss these interesting systems further in a moment. DLAs in the
merger-based models exhibit metallicity distributions very similar to the 
$f_j=1$ models, and so are not shown. 
Both the GKj25 and the BRj25 models are a good fit to the observed 
metallicity distribution in both the average metallicity and width.
We note that at $z>3$, our models begin to miss a substantial fraction 
of DLAs, which likely have lower metallicities on average. This would 
have the effect of skewing the observed distribution to lower metallicities.

The population of very low metallicity, \HI-rich, nearly ``pristine''
galaxies predicted by the metallicity-based, fixed-UV GK models is
interesting. Scaling the UV radiation field by the galaxy's
  star formation rate increases the H$_2$ fraction in low mass
  galaxies relative to the model with a UV field fixed to the MW
  value, allowing these galaxies to form significant stellar
  components.  In the fixed-UV models, the ``pristine'' galaxies are
hosted by low-mass haloes (log (M$_h /\msun) \lesssim 10$) and have
stellar masses below log M$_\ast /\msun < 6.5$.  A feature of the
metallicity-based picture for \Htwo\ formation is that if gas is low
metallicity and low density, \Htwo\ formation is extremely
inefficient, the galaxy forms few stars and the gas never becomes
enriched, so star formation stalls out. Star formation can be
``kick-started'' --- if the galaxy manages to form even a small amount
of stars, e.g. through a merger-triggered burst, this enriches the gas
leading to more star formation and enrichment, and the galaxies rather
quickly become enriched to significant levels --- hence the double
peaked distribution. The population of ``pristine'' haloes is more
prevalent in the GKfj25 model because the extended gas configuration
leads to more low surface density gas. A similar population has
recently been reported in numerical hydrodynamic simulations using a
similar metallicity-based prescription for \Htwo\ formation
\citep{kuhlen2013}. It is interesting that DLAs are observed down to
log Z = -2.5, but no DLAs have been conclusively shown to have
metallicities as low as the ``pristine'' haloes in the fixed-UV GK
models, although they could have been detected if they existed.  The
presence of a DLA metallicity floor of log Z $\gtrsim$ -2.6 has been
discussed by \cite{Wolfe2005} and \cite{Rafelski2012} while systems
with lower metallicities have been observed in the Ly$\alpha$ forest
\citep{Schaye2003, Simcoe2004}.  \citet{Qian2003} model star formation
in DLAs with a chemical evolution code and find that star formation in
pristine gas enriches it quickly, making the probability of detecting
a DLA with log Z $\lesssim$ -2.6 very small.  Additionally, we
consider whether these pockets of ``pristine'' gas could really be
common at intermediate redshifts, or whether this prediction perhaps
reflects limitations in our understanding of how \Htwo\ formation and
star formation take place in these environments.  We also note that
\cite{Gnedin2010} report that their fitting formulae, used in our GK
models, may become unreliable below ${\rm log}~ Z \lesssim -2.5$.

\begin{figure}
\includegraphics[width=3.4in]{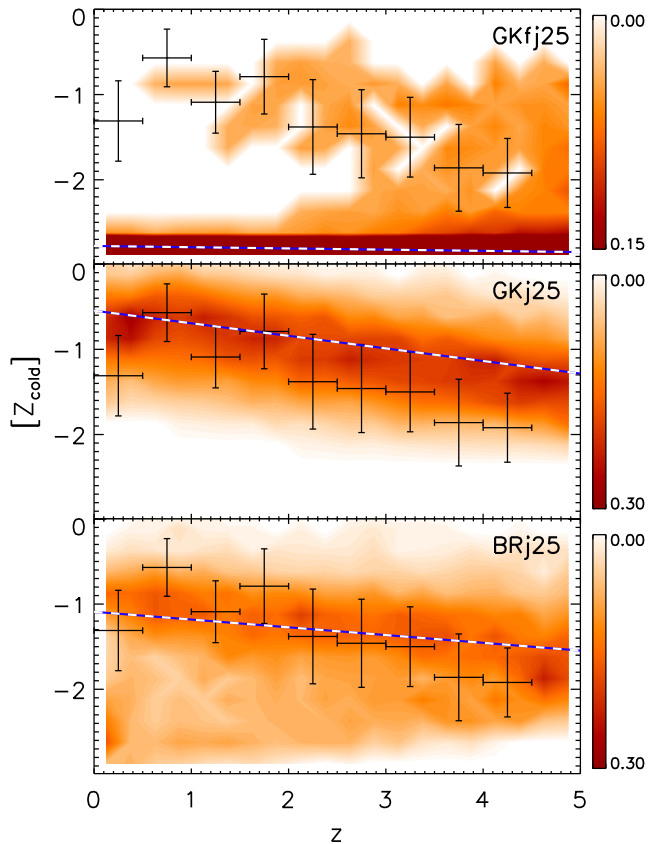}
\caption{Conditional probability of DLA metallicity as a function of
  redshift in the GKfj25 (top), GKj25 (middle), and BRj25
    models (bottom). The best linear fits to the average
  metallicities for DLAs with ${\rm log} ~ {\rm Z} > -2.5$ (to exclude
  the very low metallicity population) are overplotted (blue and
  white). Observation of DLAs from \citet{Rafelski2012} with the $1
  \sigma$ scatter in DLA metallicity are also overplotted. The
    GKj25 and BRj25 models show a shallow decrease in metallicity with
    increasing redshift, similar to observations, although the
    normalization is higher by $\sim 0.5$ dex in the GKj25 model. A
  significant number of DLAs in the GKfj25 model have metallicities
  near the ``pre-enriched'' metallicity, indicating that essentially
  no star formation has occured in these objects.}
\label{Fig:Z_z}
\end{figure}

Figure \ref{Fig:Z_z} shows the probability of selecting a DLA with a
given metallicity as a function of redshift for DLAs in the 
GKfj25 (top), GKj25 (middle), and BRj25 (top) models compared with
observational estimates from \citet{Rafelski2012}. We also show the
best linear fits to DLA metallicity as a function of redshift for each
model ([Z] $= \alpha z - [Z0]$).  In order to examine the trends with
redshift, we exclude DLAs with metallicities below log Z $ < -2.5$
from the fits.  Although all of our models predict a shallower
evolution in $Z$ with redshift than seen in by \cite{Rafelski2012}, the
agreement with observations is considerably better than in many
previous studies. Moreover, our results are also in agreement with 
\cite{Jorgenson2013} who found evidence for a lack of metallicity 
evolution over the redshift range $2.2<z<4.4$.  
We present our best-fit parameters and those from \cite{Rafelski2012} 
and \cite{Jorgenson2013} in Table \ref{Tab:Z_z}. 
We can see that the population of ``pristine'' DLAs
predicted by the GKfj25 model is present at all redshifts, and the
distribution of metallicities in the `enriched' population is as broad as 
that in the BR models. From this plot, it is apparent that the
metallicities of DLAs in our models are in better agreement with the
observational estimates in the lower part of the redshift range,
$z\lesssim 2$.

\begin{table}
\centering
\caption{DLA Metallicity vs. redshift Best-Fit Parameters}
\begin{tabular}{l c c c c}
\hline \hline
 & $\alpha_{\small \frac{dZ}{dr} = 0}$ & $ [Z0]_{\small \frac{dZ}{dr} = 0}$ & $\alpha_{\small \frac{dZ}{dr} = -0.1}$ & $ [Z0]_{\small \frac{dZ}{dr} = -0.1}$  \\
 \hline
GKfj1  & -0.10 & -0.61  & -0.07 & -0.93  \\
GKfj25 & -0.11 & -1.03  & -0.01 & -1.55  \\ 
GKj1   & -0.11 & -0.64  & -0.09 & -0.91  \\
GKj25  & -0.15 & -0.55  & -0.19 & -0.60  \\ 
BRj1   & -0.14 & -0.43  & -0.12 & -0.71  \\
BRj25  & -0.09 & -1.10  & -0.08 & -1.32  \\ 
R12$^\ast$ & -0.22 & -0.65 & - & - \\
J13$^\ast$ & -0.04 & -1.06 & - & - \\
\hline
\end{tabular}
\\
Linear fits ([Z] = $\alpha z + {\rm [Z0]}$) for DLAs at $z<5$ \\
$^\ast$ observations of DLAs at $z<5$ from \cite{Rafelski2012} and $2.2<z<4.4$ from \cite{Jorgenson2013} 
\label{Tab:Z_z}
\end{table}

Results from integral field unit spectroscopy
\citep[e.g.][]{Forster2006} indicate that star-forming galaxies at $z
\sim 2$ exhibit typical metallicity gradients of $dZ/dr =$ 0 to -0.3
dex kpc$^{-1}$ \citep[][]{Jones2010, Pilkington2012, Swinbank2012} and
up to $dZ/dr \sim -0.4$ dex kpc$^{-1}$ in quasars \citep[][]
{Jones2013}. If DLAs sample the outskirts of galaxies, e.g. with
impact parameters of b $\sim 10$ kpc at $2<z<3.5$ (typical in our
extended disc models), a metallicity gradient could have a significant
effect on typical DLA metallicities.  Moreover, metallicity gradients
in star-forming galaxies are typically measured along the disc while
DLA column density gas likely also samples a significant amount of
cold gas \emph{above} the disc where metallicities are almost
certainly lower.  Recently, \citet{Fu2013} studied the effects
  of metallicity gradients in semi-analytic models and found they
  depend on the fraction of metals directly injected into the halo as
  well as recent mergers, which can re-establish a metallicity
  gradient.  Following the observational studies mentioned above, we
implement a metallicity gradient of ${\rm dZ/dr} = -0.1$ dex
kpc$^{-1}$ where the average galaxy metallicity is set to the radius
of the average cold gas mass ($r_{ave} = 1.678 r_{g}$). Thus, galaxies
with larger cold gas masses will have more extended cold gas
distributions. In general, DLAs arising in these galaxies will have
larger impact parameters and subsequently be more affected by a
metallicity gradient. As H$_2$ formation in the GK models is
  metallicity-dependent, if included self consistently, metallicities
  gradients could also affect other galaxy properties, and might
  produce results that are different in detail from those shown
  here. We show the results of the post-processed gradients only to
  qualitatively illustrate how they can potentially affect the
  distribution of DLA metallicities. However in the BR models, H$_2$
  forms based on the midplane pressure, so introducing metallicity
  gradients would not impact other galaxy properties.

\begin{figure}
\includegraphics[width=3.4in]{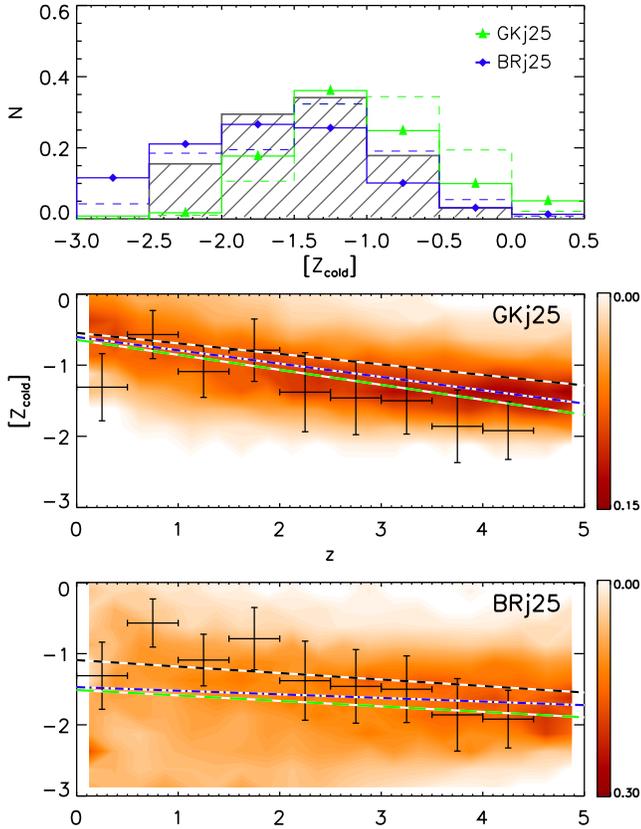}
\caption{Top panel - predicted distribution of metallicities for DLAs
  at $2< z < 3.5$ with a metallicity gradient ($dZ/dr=-0.1$ dex
  kpc$^{-1}$, solid) and without one (dashed). We exclude the fixed-UV
  and $f_j=1$ models for clarity as they exhibit similar shifts in
  metallicity. Observations of DLAs from \citet{Rafelski2012} are also
  shown in gray. Middle and bottom panels - conditional probability of
  DLA metallicity versus redshift including a metallicity gradient as
  above, for the GKj25 and BRj25 models respectively. Best
  linear fits (${\rm log} ~ Z > -2.5$) for three metallicity gradients
  are overplotted: $dZ/dr=0$ dex kpc$^{-1}$ (black and white dashed
  line), $dZ/dr=-0.1$ dex kpc$^{-1}$ (blue and white dot-dashed line),
  and $dZ/dr=-0.2$ dex kpc$^{-1}$ (green and white long-dashed line).
  Metallicity gradients can have a significant impact on the DLA
  metallicity and its evolution.}
\label{Fig:Zgrad}
\end{figure}

The top panel of Figure \ref{Fig:Zgrad} shows the distribution of DLA
metallicities in the redshift range $2<z<3.5$ before and after
implementing a metallicity gradient. For $dZ/dr = -0.1$ dex 
kpc$^{-1}$, the average shift in DLA metallicity is $\Delta 
Z=[Z_{\rm DLA}]-[Z_{\rm gal}] \sim -0.3 \pm 0.4$ with a tail
extending to $\Delta Z < -1$.  DLAs in each of our models
have a similar average shift in metallicity.
The broadened distribution reflects the large scatter in impact
parameter, which can be as high as tens of kpc, explaining the tail to
low metallicities. Although it is possible for the outskirts of
galaxies to have metallicities below log Z$ = -3$, we set this as the
floor to the DLA metallicity as it corresponds to the metallicity of
pre-enriched gas.  The predicted mean
metallicity of DLAs in this redshift range in our models is now in
quite good agreement with the observations, and the shape of the
distribution in the BRj25 models is in good agreement with the
observations, though it does produce more systems with very low 
metallicities. The addition of a metallicity gradient causes the 
GKj25 model to move into better agreement with observations.

The middle and bottom panels of Figure \ref{Fig:Zgrad} show the
conditional probability distribution of DLA metallicities with a
metallicity gradient $dZ/dr=-0.1$ dex kpc$^{-1}$ as a function of
redshift for the GKj25 and BRj25 models. The linear fits to models with
imposed metallicity gradients of $dZ/dr=0$ dex kpc$^{-1}$, $dZ/dr
Z=-0.1$ dex kpc$^{-1}$, and $dZ/dr~= -0.2$ dex kpc$^{-1}$ are shown
for reference. Metallicity gradients cause a systematic shift in the
redshift-metallicity relation for the BRj25 model while they make the 
slope steeper in the GKj25 model. 

These trends indicate that sightlines through DLAs in the BRj25 model
are sampling similar impact parameters relative to the distribution of
cold gas at all redshifts, while those in the GKj25 model are
preferentially selecting larger impact parameters at higher redshift.
On the other hand, \cite{Fumagalli2011} and \cite{Cen2012} found cold
gas to be more extended at higher redshifts, coming from nearby
streams flowing into the galaxy.

Overall, the addition of a metallicity gradient yields DLA
  metallicities in the GKj25 model in better agreement with
  observations, where stronger metallicity gradients make the above
  effects more pronounced.  We find that a metallicity gradient of
  $dZ/dr \sim -0.3$ dex kpc$^{-1}$ would be needed to bring the GKj25
  metallicities into agreement with observations while no fixed
  metallicity gradient produces satisfactory results with the BRj25
  model. In reality, the slope of the metallicity gradient probably
varies from galaxy to galaxy and may be correlated with other galaxy
properties. However, this simple exercise highlights the importance of
understanding and accounting for metallicity gradients in modeling
absorption systems.

\cite{Pontzen2008} found DLA metallicity to be correlated with halo
mass, where the majority of low metallicity DLAs (log $Z \lesssim
-1.5$) were found in haloes with masses $M_h < 10^{9.5}$ M$_\odot$. By
comparison, our simulations do not probe this mass range and we find
the typical DLA to reside in much higher mass haloes.  We also look for
a similar trend between halo mass and metallicity by dividing our
sample roughly in half based on halo mass (M$_h < 10^{11}~ \msun$,
M$_h > 10^{11}~ \msun$), and comparing their metallicity
distributions.  For each model, DLAs with halo masses $M_h < 10^{11}$ 
\msun make up the vast majority of low metallicity systems where the
difference in metallicity for each subset ranges from -0.5 to -1.0 dex.  
Finally, \cite{Neeleman2013} find a correlation between \NHI\ and
metallicity where DLAs with higher column densities have higher
metallicities.  We also see a similar, but weak trend in our model
DLAs.  Interestingly, the addition of a metallicity gradient causes
this correlation to be \emph{stronger} as it causes sightlines that
sample the interior regions of galaxies to preferentially have higher
metallicities than those sampling the outskirts of galaxies.
Therefore, a strong correlation between metallicity and \NHI\ may also be
suggestive of the presence of metallicity gradients in DLAs. A
more detailed comparison between metallicities measured through
absorption studies and via emission lines in star-forming galaxies,
will be presented in Berry et al. (2014, in prep.).

\subsection{Kinematics}
\label{Sec:dv90}

Observationally, low ionization line profiles in DLAs exhibit multiple
components with small individual velocity dispersions, yet the
relative velocities between each component can be quite large.
Therefore, \cite{Prochaska1997} defined a number of statistical
measures to probe the characteristics of line profiles. The velocity
interval, $\Delta v_{90}$, is defined as the difference in velocity
between the pixel containing 5\% and 95\% of the total optical
depth. We calculate this statistic for each of our DLAs based on their
low ionization absorption line profiles as detailed in Section
\ref{Sec:ModelDLA}. This quantity has proven to be the most difficult
to reproduce in theoretical models, so we focus on $\Delta v_{90}$ for
our analysis.

\begin{figure}
\includegraphics[width=3.4in]{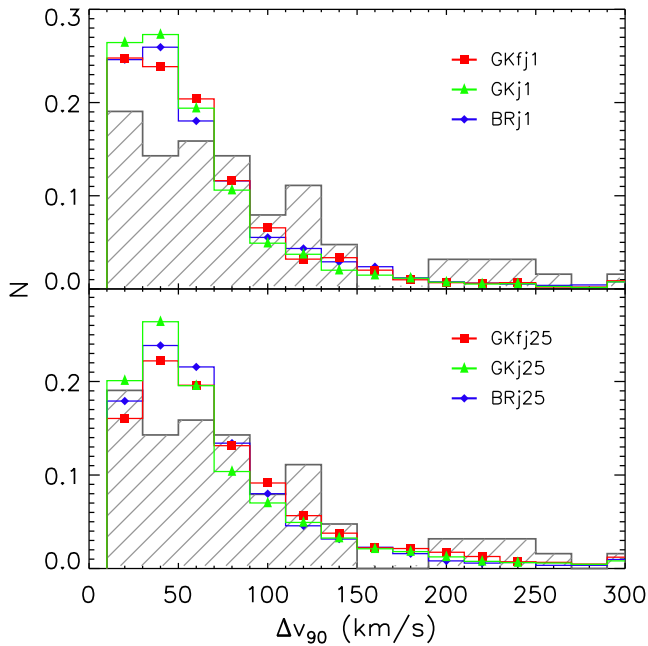}
\caption{Upper panel -- distribution of $\Delta v_{90}$ values for DLAs
  in the redshift range $2<z<3.5$ for the GKfj1 (red), GKj1
    (green), and BRj1 (blue) models. Lower panel -- same as top except
    for the GKfj25 (red), GKj25 (green), and BRj25 (blue) models. The
  observed $\Delta v_{90}$ distribution from \protect\citet{Neeleman2013} is
  overplotted for reference. The GKj25 model has a $\Delta v_{90}$
  distribution most consistent to observations of high-redshift
  DLAs. DLAs with $\Delta v_{90} > 300$ km s$^{-1}$ are shown at
  $\Delta v_{90}= 300$ km s$^{-1}$. The $f_j=1$ models produce too
  many low-$\Delta v_{90}$ DLAs as has been shown in previous analyses
  of DLAs in semi-analytic models \protect\citep[e.g.][]{Maller2001}.}
\label{Fig:dv90}
\end{figure}

Figure \ref {Fig:dv90} shows the distribution of velocity widths,
$\Delta v_{90}$, for DLAs in the $f_j=1$ and $f_j=2.5$ models
with the $\Delta v_{90}$ values measured from observed DLAs
overplotted \citep{Neeleman2013}. Historically, hierarchical models
have had difficulty reproducing this distribution, tending to
overproduce low-$\Delta v_{90}$ objects and underproduce the number of
high-$\Delta v_{90}$ systems \citep[e.g.][]{Maller2001, Pontzen2008}.
The $f_j=1$ models have $\Delta v_{90}$ distributions
that peak at low $\Delta v_{90}$, which is due to the large number of
DLAs hosted by low-mass dark matter haloes.  
On the other hand, the $f_j=2.5$ models
are noticeably better at reproducing the observed distribution of
$\Delta v_{90}$.  Table \ref{Tab:dv90} shows the probability that the
distribution of DLA $\Delta v_{90}$ values in each model are taken
from the same distribution as the observed $\Delta v_{90}$ values as
well as the D-statistic from the Kolmogoroz-Smirnov test. All $f_j=1$
models have a $\lesssim 1\%$ chance of being from the same
distribution as the observed $\Delta v_{90}$ values.  However, the
$f_j=2.5$ models are more successful, with the GKj25 model producing
the most realistic $\Delta v_{90}$ distribution with a probability of
27\%. This kinematic distribution is a significant improvement over
those from previous semi-analytic models \citep[e.g.][]{Maller2001,
  Maller2003} and many numerical simulations. Our models do not
include other effects such as galactic winds or cold accretion, which
will likely increase the $\Delta v_{90}$ value of a given DLA
\citep[e.g.][]{Cen2012}. Note that the $f_j=1$ models are in slightly
better agreement with the data than the $f_j=1$ discs from previous
semi-analytic models \citep[e.g. see Figure 1 in][]{Maller2001}.

\begin{table}
\centering
\caption{DLA $\Delta v_{90}$ K-S Test}
\begin{tabular}{l c c}
\hline \hline
 & D & P$_{\rm K-S}$  \\
 \hline
GKfj1  & 0.18 &  2.5e-4 \\
GKfj25 & 0.14 & 0.017  \\ 
GKj1   & 0.23 &  2.5e-3 \\
GKj25  & 0.12 & 0.27  \\ 
BRj1   & 0.23 & 2.6e-3 \\
BRj25  & 0.18 & 0.035  \\ 
\hline
\end{tabular}
\\
For DLAs at $2<z<3.5$, compared to observations from \cite{Neeleman2013}.
\label{Tab:dv90}
\end{table}

The improvement in the $\Delta v_{90}$ distribution in the $f_j=2.5$ models 
is primarily due to two effects. Most importantly, more extended cold
gas discs cause DLAs to originate in more massive haloes with larger
rotational velocities, as we saw in Figure~\ref{Fig:cs_Mh}.  As a secondary
effect, more extended gas distributions make it more likely for DLAs
to arise from `multiple hits' or lines of sight passing through more
than one galaxy within a parent halo.  The large number of low mass
galaxies in the GKfj25 model causes an excess of DLAs with low
$\Delta v_{90}$ values. The improved success of the $f_j=1$ models is due
to the increased cold gas masses of DLAs in all models causing more
higher mass DLAs to be selected. 
We note that when cold gas discs are made very thin (e.g. very small 
values of $\chi_z$), the $\Delta v_{90}$ distributions also become 
peaked at low values.

\begin{figure*}
\includegraphics[width=6in]{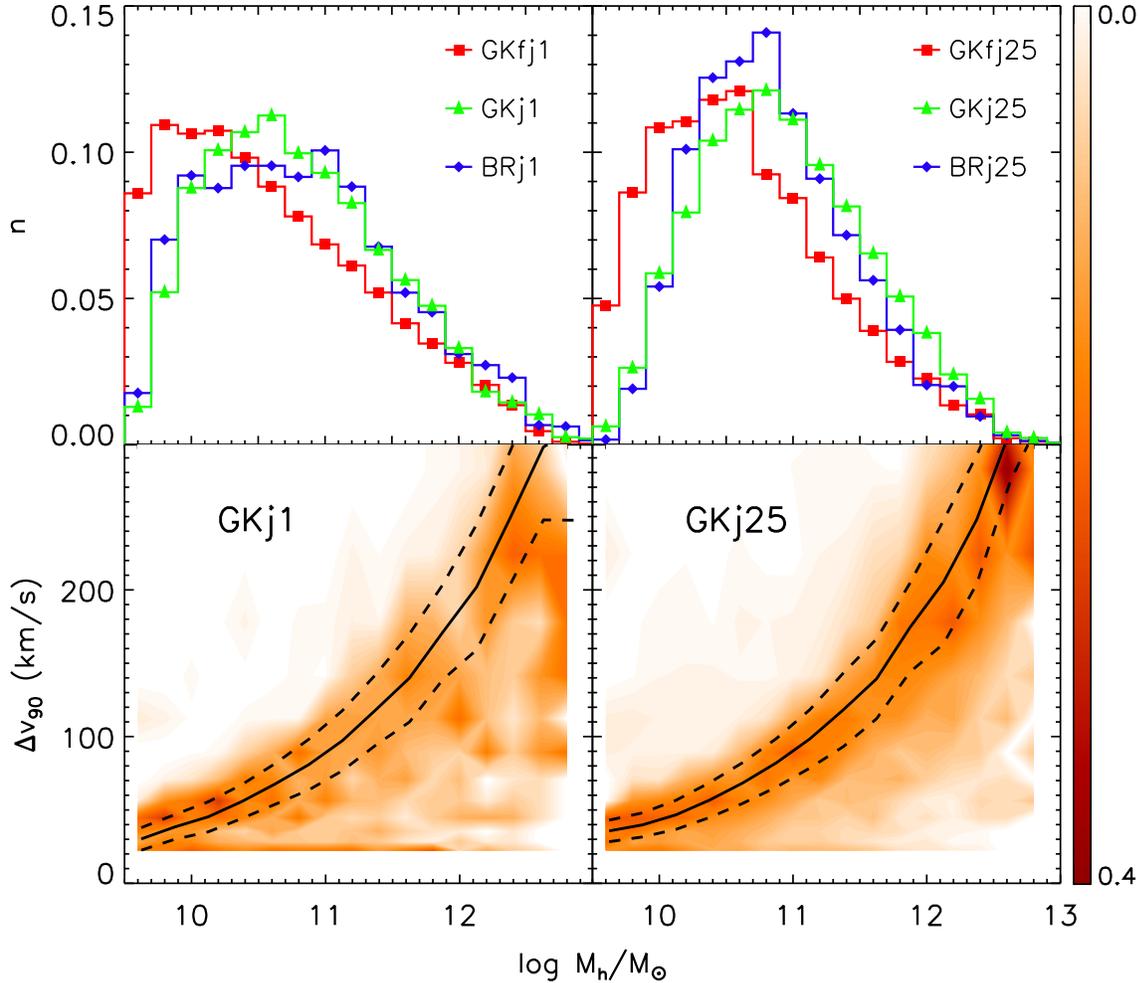}
\caption{Upper left - histogram of the fraction of DLAs with a
    given halo mass for the GKfj1 (red), GKj1 (green), and BRj1 (blue)
    models at $2<z<3.5$. Upper right - same as upper left except for
    the GKfj25 (red), GKj25 (green), and BRj25 (blue) models. Lower
    left - conditional probability of $\Delta v_{90}$ versus DLA halo
    mass for the GKj1 model. Lower right - same as lower left except
    for the GKj25 model. The BRj1 and BRj25 models produce similar
    $\Delta v_{90}$-halo mass relations as the GKj1 and GKj25 models
    respectively. The average disc velocity (solid line) and
  $1\sigma$ scatter (dashed lines) are overplotted for reference. The
  higher halo masses of DLAs in the $f_j=2.5$ models cause them to
  produce more high-$\Delta v_{90}$ systems. DLAs with low halo mass
  and high $\Delta v_{90}$ values seen in the bottom row arise from
  `multiple hits' or lines of sight passing through more than one
  galaxy in the same halo. }
\label{Fig:Mh_dv90}
\end{figure*}

The upper row of Figure \ref{Fig:Mh_dv90} shows the distribution of
halo masses for DLAs in our models over the redshift interval
$2<z<3.5$.  The lower row shows the probability of
finding a DLA with a given $\Delta v_{90}$ value at a given halo mass at the
same redshift. The average disc circular velocity and $1\sigma$
scatter for galaxies at a given halo mass are also overplotted.  We
find a similar trend between $\Delta v_{90}$ and disc velocity as
\cite{Haehnelt1998} who found that $\Delta v_{90}$ tracked disc
velocity following $\Delta v_{90} \sim 0.6 v_{\rm disc}$ with a large
scatter.  As can be seen in the top and bottom panels of Figure
\ref{Fig:Mh_dv90}, the extended gas distributions cause DLAs in the
$f_j=2.5$ models relative to the $f_j=1$ models, specifically the 
GKj25 and BRj25 models, to originate from more massive haloes, which in 
turn have higher disc velocities.  Conversely, the large number of pristine 
systems in very low mass haloes in the fixed-UV GK models lowers the average 
halo mass and $\Delta v_{90}$ value relative to the other models.  DLAs
with large $\Delta v_{90}$ values and low halo masses typically arise
from `multiple hits', which are more common in the $f_j=2.5$ models as
can be seen in the bottom row of Figure \ref{Fig:Mh_dv90}.

\begin{figure}
\includegraphics[width=3.4in]{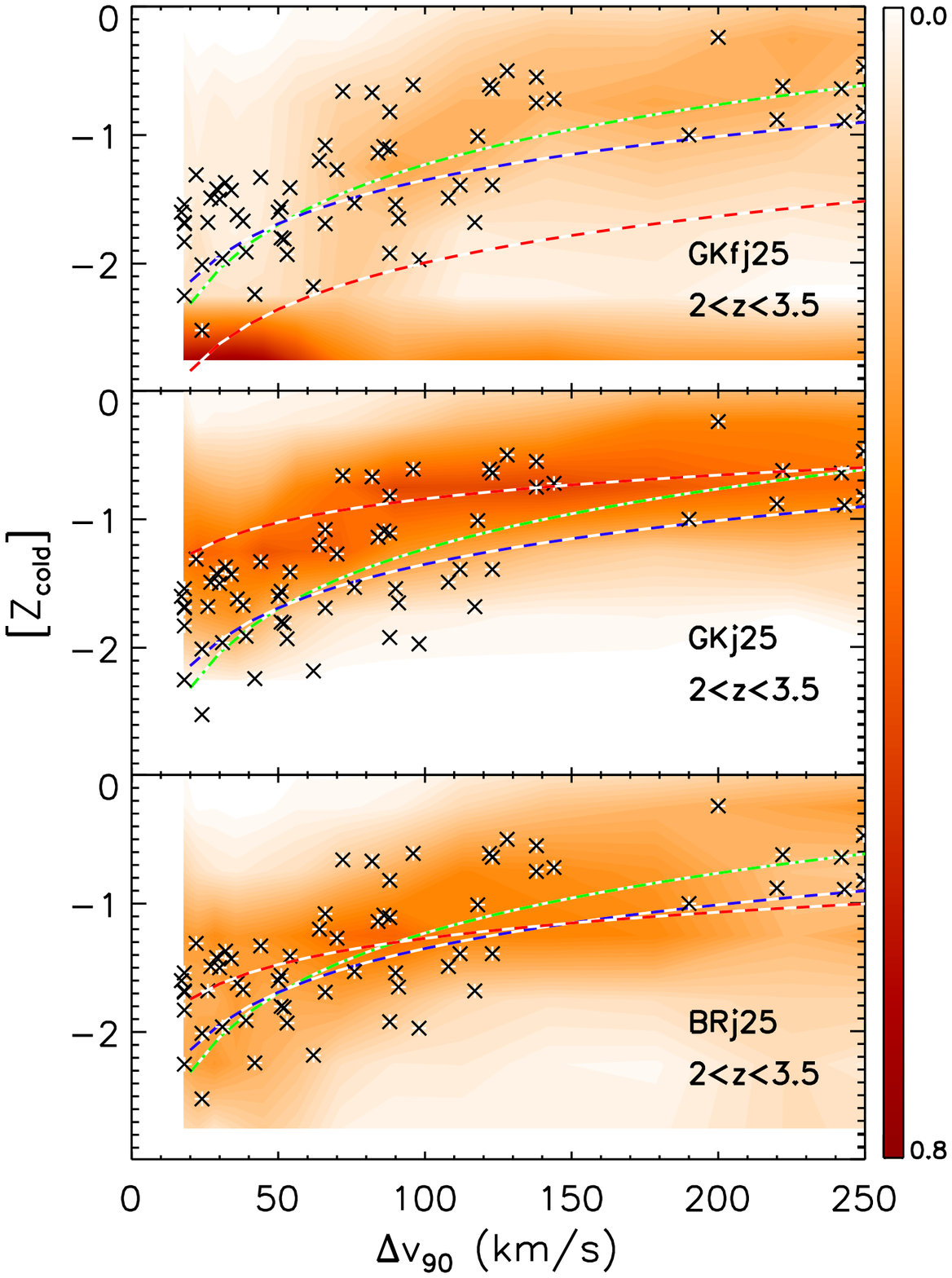}
\caption{Conditional probability of $\Delta v_{90}$ versus metallicity
  for DLAs in the redshift range $2<z<3.5$. Top panel - GKfj25
    model. Middle panel - GKj25 model. Bottom panel - BRj25 model. The
    best-fit power law for each model is shown by the red and white
    dashed line. We overplot the observed DLA $\Delta v_{90}$ versus
    metallicity relation from \citet[][crosses]{Wolfe2005},
    \citet[][green dot-dashed line]{Ledoux2006}, and \citet[][blue
      dashed line]{Jorgenson2013}. All models show evidence for an
    increase in metallicity with $\Delta v_{90}$, where the BRj25 model
    is in the best agreement with observations while the GKj25 model
    does not produce any very low-Z DLAs. At high $\Delta v_{90}$, the
    BRj25 model produces DLAs with metallicities in good agreement
    with observations. }
\label{Fig:Z_dv90}
\end{figure}

Figure \ref{Fig:Z_dv90} shows the conditional probability distribution
of DLA metallicity as a function of $\Delta v_{90}$ for the
GKfj25, GKj25, and BRj25 models. The best-fit power law to 
the metallicity-$\Delta v_{90}$ relation for our models and observations 
is also shown for reference, and the best-fit parameters are given in Table 
\ref{Tab:Z_dv90}. In the $f_j=2.5$ models, systems with high
$\Delta v_{90}$ values have metallicities comparable to observations,
while they are slightly higher in the $f_j=1$ models. At low
$\Delta v_{90}$, DLAs in the BRj25 model are in reasonable agreement with 
observations while those in the GKj25 model typically have slightly
larger metallicities than is observed.  Alternatively, the GKfj25 
model produces a significant number of low metallicity, low $\Delta
v_{90}$ systems, but again the metallicity distribution of these objects 
is peaked towards lower values in the models than the observations
indicate. DLAs in the $f_j=1$ models are not shown as they predict 
metallicities and $\Delta v_{90}$ values that are significantly higher 
and lower, respectively, than observations.

Numerous authors have proposed that the $\Delta v_{90}$-metallicity
relation may be the DLA version of the well-known mass-metallicity
relation \citep[e.g.][and references therein]{Prochaska2008,
  Moller2013, Neeleman2013}.  If the motions of neutral clouds giving
rise to the $\Delta v_{90}$ values are governed by gravity whether it
is rotation, infall, or outflow, then $\Delta v_{90}$ could be
correlated with mass. Indeed, as shown in Figure \ref{Fig:Mh_dv90},
our models do predict a relationship between $\Delta v_{90}$ and halo
mass, although with significant scatter. Similarly, we find galaxies
with large $\Delta v_{90}$ values to be more likely to have higher
metallicities. At all values of $\Delta v_{90}$, there is significant
scatter in DLA metallicity, which is likely affected by galaxy
inclination, the stochasticity of neutral clump properties, and
infalling or outflowing gas.  Moreover, we do find that the presence
of a metallicity gradient in all DLAs has the effect of increasing the
scatter of the $[Z] - \Delta v_{90}$ distribution as the high-$\Delta
v_{90}$ DLAs are selected from higher mass haloes with a larger range
of impact parameters. For this reason, we note that a metallicity
gradient can weaken the connection between metallicity and $\Delta
v_{90}$. Observations of star-forming galaxies show a large range in
metallicity gradients, suggesting that this may also be the case in
DLAs. Overall, metallicity gradients may contribute to the low
observed metallicities and large scatter seen in DLAs.

\begin{table}
\centering
\caption{DLA Metallicity vs. $\Delta v_{90}$ Best-Fit Parameters}
\begin{tabular}{l c c}
\hline \hline
 & $\alpha_{\small dZ/dr = 0}$ & $ [Z0]_{\small dZ/dr = 0}$ \\
 \hline
GKfj1  & 0.66 & -2.1   \\
GKfj25 & 0.77 & -2.9   \\ 
GKj1   & 0.66 & -2.1   \\
GKj25  & 0.61 & -2.1   \\ 
BRj1   & 0.69 & -2.0   \\
BRj25  & 0.68 & -2.6   \\ 
L06$^\ast$ & 1.55 & -4.33 \\
J13$^\ast$ & 1.13 & -3.61 \\
\hline
\end{tabular}
\\
Linear fits ([Z] = $\alpha {\mathrm log} \Delta v_{90} + {\rm [Z0]}$) for DLAs at $2<z<3.5$. \\
$^\ast$ observations of 70 absorbers with log N$_{HI} > 20$ at $1.7<z<4.3$ from \cite{Ledoux2006} and 106 DLAs at $2.2<z<4.4$ and \cite{Jorgenson2013}.
\label{Tab:Z_dv90}
\end{table}

\section{Discussion}
\label{Sec:Discussion}

Studies of HI in absorption are currently the \emph{only} means of
probing atomic gas at significant redshift, and this will remain true
for quite some time at redshifts much greater than unity, although new
radio telescopes such as the Square Kilometer Array (SKA) and its
precursors will push studies of \HI\ in emission up to $z\sim
1$--1.5. DLAs are thought to contain the bulk of the \HI\ in the
Universe and extensive observations of their properties exist in the
literature. In spite of this, theoretical models of galaxy formation
based on the predominant \LCDM\ paradigm have not been terribly
successful at reproducing several of these observations, and therefore
the connection between DLAs and the galaxy population detected through
their emission properties remains unclear.

Although our models rely on a number of simplifying
assumptions, our study is the first to attempt to predict
the properties of DLAs in a model that includes both the
cosmological formation of galaxies and the partitioning of the cold
ISM in galaxies into different phases (ionized, atomic, and
molecular), based on physically motivated recipes. In addition, unlike
the numerical hydrodynamic simulations that have been predominantly
used for previous studies of DLA properties, our semi-analytic models
broadly reproduce a large suite of observations that probe the stellar
and dust content of galaxies over a broad redshift range (S08, S12).

\subsection{On the Distribution of Cold Gas giving rise to DLAs}
\label{Sec:Disc_gas}

An interesting general insight is that \emph{in spite} of the freedom
we allowed ourself in adjusting, for example, the distribution of cold
gas in galactic discs, \emph{none} of the models we considered is
really fully satisfactory at reproducing the properties of DLAs over
all redshifts \emph{and} the set of `calibration' quantities of $z=0$
galaxies (e.g. stellar mass functions, HI mass functions, \Htwo\ mass
functions). We found that models with ``standard'' distributions of
gas ($f_j=1$) could not reproduce the column density distribution of
DLAs at any redshift, in agreement with the results of previous
studies \citep{Maller2001}. This led us to consider models with
`extended' gas distributions, which reproduced the HI column density
distribution, the cosmic evolution of the line density, and
$\Omega_{\rm DLA}$ quite well at $z \lesssim 3$. However, the models
with extended gas distributions and a metallicity-dependent recipe for
\Htwo\ formation were too inefficient at forming stars and predicted
stellar fractions that were too low, and HI fractions that were too
high, at $z=0$. This is because the extended gas models place a large
amount of gas at low surface density where it is inefficient at
forming stars. The metallicity-based extended-gas model with a
  varying-UV background (GKj25) performs better than the model with a
  fixed UV background, as the reduced UV background in low-mass
  systems results in weaker suppression of the star formation
  efficiency.
The pressure-based extended-gas BRj25 recipe fared better with
the $z=0$ stellar mass function, but still overproduces the
number of \HI-rich galaxies at $z=0$. 

This may reflect the overly simple assumption of a fixed value of
$f_j$ in all galaxies and at all redshifts that we made in this
work. We motivated our assumption based on numerical hydrodynamic
simulations. These simulations have shown that the average specific
angular momentum $j$ of the baryonic disc can be larger than that of
the DM halo if low-$j$ material is preferentially ejected by winds
\citep[e.g.][]{Brooks2011, Guedes2011}, if the disc is formed by cold
flows \citep{Brooks2011, Pichon2011, Stewart2013}, or if the gas is
spun up in a gas-rich major merger \citep{Robertson2006, Sharma2012}.
It is likely that the effective value of $f_j$ in reality varies
between galaxies and may depend on cosmic epoch and the star formation
and merger history. Our simple ``merger'' models did not fare well in
reproducing the column density distribution or the $\Omega_{\rm DLA}$
evolution, indicating that a more physical approach to modeling the
extent of the cold gas giving rise to DLAs is needed.

In addition, some authors have suggested that some of the absorbing
gas giving rise to DLAs may not be in a rotationally supported disc at
all, but may instead be in an outflow \citep{Cen2012}. Although we
stated earlier that our `extended' disc models could be interpreted as
representing any of these scenarios, this is not entirely
consistent. Our models implicitly assume that DLAs arise from the cold
gas reservoir that has accumulated within the ``box'' that we call a
galaxy -- i.e., gas that has cooled and accreted but not turned into
stars or been ejected in an outflow. Assuming that this gas is in an
extended configuration leads to large amounts of gas at low surface
density where it is inefficient at forming stars. If the cross-section
and kinematics of DLAs are significantly affected by gas in an
outflow, this would have different implications for the star formation
efficiency of galaxies which in turn feeds back into the cold gas
content as well as metallicities, etc. We are engaged in a program to
study the distribution of gas and metals in high resolution zoom
simulations that include a treatment of a multi-phase ISM
\citep{Christensen2012} in order to study these issues and build more
sophisticated semi-analytic models for absorption systems.

\subsection{Is the flatness of $\Omega_{\rm DLA}$ a Cosmic Coincidence?}

Regardless of these limitations, we believe that we can make some
robust and interesting predictions based on our simple
models. Our preferred models, the BR/varying-UV GK models with
  extended discs, are somewhat successful at reproducing the line
  density of DLAs and $\Omega_{\rm DLA}$ at $z\lesssim 3$, indicating
  that the majority of HI in DLAs could plausibly be associated with
  galactic discs.  Subsequently, one interesting suggestion we make
is that the observed near constancy of $\Omega_{\rm DLA}$ from $z\sim
4.5$ to $0.5$ is not a reflection of a truly constant global
\HI\ density, but rather a cosmic coincidence resulting from the DLA
population being increasingly `contaminated' by intergalactic gas in
filaments or cold streams at $z\gtrsim 3$, and systematically
under-representing the true \HI\ density at $z\lesssim 2$ due to an
increasing contribution to the global \HI\ budget from
low-\NHI\ systems.  Numerical hydrodynamic simulations do indicate
that the number of ``intergalactic'' DLAs increases with increasing
redshift \citep{Fumagalli2011, Cen2012}.

At the same time, it is a concern that in several of our models 
(in particular the overall more successful BR/varying-UV GK models) 
the \emph{total} cold gas density in all forms (atomic, and molecular) 
barely equals (BRj25/GKj25/GKfj1) or is even less than (BRj1/GKj1) the \HI\ 
density implied by observations of DLAs at $z\gtrsim 3$. On the other hand 
in the fixed-UV, extended gas model (GKfj25), the total amount of stars at $z=0$
is much less than observations. 
We note that a similar problem
is seen in many other semi-analytic models and numerical hydrodynamic
simulations \citep[e.g.][]{Duffy2012, Narayanan2012, Dave2013}. This
may be a reflection of another well-known problem, that low-mass
galaxies in these models form too early \citep{Fontanot2009,
  Weinmann2012}. Thus overly efficient star formation at high redshift
may be consuming or expelling too much gas, leaving behind too little
cold gas in galaxies. We note that the models with metallicity-based
\Htwo\ formation recipes do appear to fare better in this regard.
Moreover, this effect may also be impacted by the decreasing average
halo mass of DLAs with redshift and the mass resolution of our models.
In this case, a contribution from low mass DLAs below our resolution
limit may explain part of this discrepancy.

\subsection{DLA Host Masses and Kinematics}

Correlation studies of DLAs are an interesting and powerful method to
directly constrain the DLA cross section as a function of halo mass.
Through a cross correlation analysis of DLAs with the Ly$\alpha$
forest at $\langle z\rangle=2.3$, \cite{Font-Ribera2012} recently
found typical DLA halo masses of $M_h \sim 6 \times 10^{11}~ \msun $
and DLA cross sections $\sDLA\ \sim 1400$ kpc$^{2}$. Our
  `extended disc' ($f_j=2.5$) models, specifically the BRj25 and GKj25
  models, produce DLAs with comparable cross sections and halo masses
  at the same redshift, and the predicted slope of the DLAs
  cross-section vs halo mass ($\sDLA$-M$_h$) is in reasonable
  agreement with these observations.

We find that the distribution of $\Delta v_{90}$ for our model DLAs
depends on both the gas partitioning recipe and the specific angular
momentum of the gas ($f_j$). The $\Delta v_{90}$ distributions for our
`extended disc' models are significantly better than results from
previous SAMs \citep[e.g.][] {Maller2001} and most numerical
hydrodynamic simulations, with our metallicity-based extended-gas, varying-UV 
(GKj25) model providing the closest match.  Similarly to \cite{Haehnelt1998},
we find $\Delta v_{90}$ and disc rotation velocity to be positively
correlated, although with a large scatter. 
The disc rotation velocity corrected for inclination is the primary factor 
determining the value of $\Delta v_{90}$.
We show that adopting the extended gas discs shifts DLAs into more massive 
haloes, which leads to better agreement with the observed $\Delta v_{90}$
distribution. We still predict fewer very large $\Delta v_{90}$
systems ($\Delta v_{90}\gtrsim 200$ km/s) than are observed. This may
be because $\Delta v_{90}$ is boosted by outflows in some systems, as
suggested by the simulations of \cite{Cen2012}.
Additionally, outflows are known to be common among high-redshift 
star-forming galaxies \citep[e.g.][and references therein]{Steidel2010}. 

\subsection{DLA Metallicities}
 
Another insight from our study is a first indication of how
observations of gas at high redshift may help constrain our
understanding of how \Htwo\ forms in galactic discs and how cold gas
is converted into stars on galactic scales. Some workers have recently
emphasized the importance of metallicity in regulating
\Htwo\ formation and hence star formation \citep[e.g.][]{Krumholz2009,
  Gnedin2010}. However, other authors have suggested that the ISM is
self-regulated on the scales of star-forming regions, so that the star
formation rate adjusts itself until the turbulent pressure balances
the gravitational restoring pressure, thus suggesting that the disc
midplane pressure may be the more fundamental quantity
\citep{Ostriker2010, Shetty2012}. Although the
  metallicity-based, varying-UV background GK model produces slightly
  better results for DLAS, perhaps the more striking result is how
  similar are these results overall to those of the pressure-based BR
  model. In addition, we find that including the effects of a varying
  UV background in the metallicity-based model produces significantly
  improved results relative to a model in which the galaxy-to-galaxy
  variation in the UV field is neglected \citep[e.g.][]{Krumholz2009}.
  The fixed-UV GK models make the dramatic prediction that a large
  fraction of DLAs at all redshifts are composed of nearly
  ``pristine'' gas with metallicity log $Z \lesssim -2.5$, that has
  never experienced significant star formation.  These arise because
the metallicity-based models predict that low surface density,
low-metallicity gas is extremely inefficient at forming \Htwo. In our
models, stars can only form out of \Htwo\ so in these systems, stars
never form efficiently and never enrich the gas, so remain ``barren''
unless kick-started by a merger-driven starburst. We were initially
concerned that these systems were an artifact of inaccuracies in our
models, but a similar population has recently been pointed out in
numerical hydrodynamic simulations \citep{kuhlen2013}, which
incorporate a similar metallicity-based recipe for \Htwo\ formation
and \Htwo-based star formation. Therefore it is interesting to
determine how robust these predictions are, and whether such a
population is definitively ruled out by observations.

\cite{Wolfe2005} and \cite{Rafelski2012} discuss the presence of a DLA
metallicity floor of log Z $\gtrsim -2.6$, and no lower metallicity
DLAs have been discovered, although they could have been
detected. \cite{Schaye2003} and \cite{Simcoe2004} find systems in the
Ly$\alpha$ forest with lower metallicities, down to log Z $\sim
-4$. It is worth noting in this context, however, that we assume that
the fitting functions based on numerical simulations with chemistry
and simplified radiative transfer hold down to gas metallicities of
$\log Z \sim -3$, while \citet{Gnedin2010} note that these
approximations are uncertain below metallicities of $\log Z \sim-2.5$.
In addition, recent high resolution numerical simulations suggest that
in low metallicity gas, \HI\ gas can form directly into stars without
going through a molecular phase \citep{Shetty2012, Glover2012,
  Shetty2013}.

Another success of our models is improved agreement with the low
metallicities and weak metallicity evolution of DLAs, which has been a
challenge for previous generations of SAMs and numerical simulations.
The reduced efficiency of star formation in systems with low mass
and/or low metallicity causes DLAS to typically have lower
metallicities than in previous SAMs, thereby increasing the slope of
the mass-metallicity relation.  
Our extended-gas BRj25 and GKj25 models, with
  the possibility of observationally motivated metallicity gradients,
  reproduce the mean metallicities of DLAs at $z\lesssim 3$, including
  the slope of the redshift evolution. At $z\gtrsim 3$, the varying-UV
  GK model predicts metallicities higher than the observations by up
  to 0.5 dex while the pressure-based BR model is in better
  agreement. It is worth noting that as we discussed above, the
  observed population of DLAs in this regime may be significantly
  contaminated by intergalactic gas which would be expected to have
  lower metallicity. 
However, more recently, \cite{Jorgenson2013} find a much weaker
evolution in DLA metallicity than \cite{Rafelski2012}. Their sample
has a more limited redshift range of $2.2<z<4.4$, and they note that
they find much flatter metallicity evolution when they take the data
set from \cite{Rafelski2012} and only include DLAs in the same
redshift range.  These results suggest that a significant amount of
evolution in the DLA metallicity occurs at $0<z<1$ and $z>4$
\citep[see figure 11 in][]{Jorgenson2013}.

Metallicity gradients as steep as $dZ/dr = -0.3$ dex kpc$^{-1}$
have been observed in $z \sim 2-2.5$ star-forming galaxies
\citep{Jones2010}. We showed that introducing metallicity gradients of
this order significantly changed the predicted metallicity
distribution of DLAs and its redshift evolution. Thus our work shows
that metallicity gradients may play a significant role in
reconciling DLAs metallicities and their evolution with those measured
via emission line diagnostics, underlining the importance of
understanding and constraining these gradients in simulations and
observations.

Another possibility that has been discussed in the literature is that
the metallicities of DLAs could be biased low due to a selection
effect, as metal-rich DLAs would also be dusty, perhaps causing their
background quasars to drop out of colour or magnitude selected quasar 
samples.  However, using radio-selected quasars, \citet{Ellison2001}
and \citet{Khare2012} found no significant population of heavily
reddened quasars with foreground DLAs, suggesting that DLAs do not
strongly redden quasars.  Furthermore through stacking quasars with
and without DLAs along the line of sight, \citet{Ellison2005} found
that DLAs are associated with very small amounts of dust reddening.
These analyses suggest that observations of DLAs are not missing
dusty, evolved DLAs and that they are well-characterized by small
amounts of dust and low metallicities.

\section{Conclusions}
\label{Sec:Conclusions}

We have presented predictions for the properties of quasar absorption
line systems associated with cold gas in galactic discs, based on a
new suite of semi-analytic models that include updated recipes for
partitioning gas into an ionized, atomic, and molecular phase, and a
molecular hydrogen based star formation recipe. We present
  results for three different prescriptions for partitioning gas into
  an atomic and molecular component: one (BR) based on the empirical
  relationship between molecular fraction and gas midplane pressure
  from \citet{Blitz2006}, and two (GK) based on numerical hydrodynamic
  simulations in which the metallicity and the local UV radiation
  field play a major role in determining the molecular fraction
  \citep{Gnedin2010,Gnedin2011}. In one, the local UV radiation field
  is held fixed to the Milky Way value, while it is allowed to vary in
  proportion to the galaxy star formation rate in the other. In addition,
we considered different approaches for how the cold gas is distributed
within galaxies, using the parameter $f_j$ to represent the fraction
of the specific angular momentum of the dark matter halo material that
is ``captured'' by the cold gas within galaxies.  First, we compared
our predicted $z \sim 0$ galaxy stellar mass functions and cold gas
mass functions to local observations. We find that models with
different physical ingredients produce different amounts of cold gas
within galaxies, demonstrating that it is a key diagnostic for
constraining theoretical models.  We then selected DLAs by passing
sightlines through mock catalogs extracted along lightcones from the
Bolshoi simulation. The majority of our analysis focused on comparing
the properties of these mock DLAs with observations of real DLAs.

We summarize our main results below.

\begin{itemize}

\item Models with ``standard" gas radial profiles where the cold gas
  specific angular momentum is equal to that of the dark matter,
  $f_j=1$, fail to reproduce the observed column density distribution
  function and the number density of DLAs.  They also fail to
  reproduce the observed distribution of velocity widths, $\Delta v$,
  of low-ionization metal lines, producing too many low-$\Delta v$
  systems and not enough high--$\Delta v$ systems.

\item Our models with ``extended" gas discs ($f_j=2.5$) are able to
  reproduce the observed column density distribution function over the
  range $19 < \rm{log}~ \NHI < 22.5$ at $2<z<3.5$ and also the number
  density of DLAs at $z<2.5$.  The favored extended disc
    models produce comparable DLA cross sections ($\sigma_{DLA}$),
    halo masses, and predict a $\sigma_{DLA}-\rm M_h$ relation with
    slope and normalization that is in agreement with observations at
    $\langle z\rangle=2.3$.  Moreover at $2<z<3.5$, these same models
    also reproduce the $\Delta v$ distribution fairly well, primarily
    because they select DLAs in higher mass haloes with larger
    rotational velocities. However, they produce too many
    high-M$_{HI}$ mass galaxies and too few galaxies around the
    characteristic stellar mass at $z=0$. 

\item Our ``extended" gas disc models reproduce the observed
  estimates of the comoving density of \HI\ in DLAs ($\Omega_{\rm
    DLA}$) at $z \lesssim 3$, with the model with pressure-based gas
  partitioning and the one with metallicity-based gas partitioning
  accounting for a varying UV radiation field producing the best
  agreement.  All of our models fail to reproduce the number density
  of DLAs and amount of \HI\ in DLAs at $z \gtrsim 3$. We suggest that
  a significant fraction of DLAs at $z \gtrsim 3$ may reside in
  filaments, cold streams, or clumps of gas not associated with the
  rotationally supported gas in galactic discs.

\item The predictions of our favored ``extended" gas disc
  models for DLA metallicities are in much better agreement with
  observations than previous studies. Additionally, accounting for
  metallicity gradients motivated by observations of star-forming
  galaxies at a comparable redshift to the DLA sample, we obtain very
  good agreement with the mean metallicity and metallicity-redshift
  relation in our models.  Furthermore, our models predict a correlation
  between metallicity and kinematics ($\Delta v_{90}$) which is in
  reasonable agreement with observations.

\end{itemize}

We thank Marcel Neeleman, Dusan Keres, and Danail Obreschkow for
providing their observational data in electronic form, and we thank
Art Wolfe, Andrey Kravtsov, and Eve Ostriker for enlightening
discussions. We dedicate this work to the memory of Art Wolfe. 
EG is supported by NSF Career Grant 1055919. 
AM is supported by NSF Grant 1153335.


\label{lastpage}

\end{document}